\documentclass[a4paper,12pt]{article}
\pdfoutput=1
\usepackage{epsfig}
\usepackage{amsmath}
\usepackage{amssymb}
\usepackage{amsfonts}
\usepackage{bbm}
\usepackage{fleqn}
\usepackage[footnotesize]{caption}
\usepackage{graphicx}
\usepackage{mathrsfs}
\usepackage[center,footnotesize,hang]{subfigure}

\def\be{\begin{equation}}
\def\ee{\end{equation}}
\def\beq{\begin{equation}}
\def\eeq{\end{equation}}
\def\bea{\begin{eqnarray}}
\def\eea{\end{eqnarray}}

\def\<{\left\langle}
\def\>{\right\rangle}

\def\lsim{\stackrel{<}{{}_\sim}}

\def\be{\begin{equation}}
\def\ee{\end{equation}}
\def\beq{\begin{equation}}
\def\eeq{\end{equation}}
\def\bea{\begin{eqnarray}}
\def\eea{\end{eqnarray}}

\newcommand{\newc}{\newcommand}
\newc{\ol}{\overline}
\newc{\wt}{\widetilde}
\newc{\bs}{\boldsymbol}
\newc{\ma}{\mathcal}
\newc{\vl}{\langle}
\newc{\vr}{\rangle}
\def\vev#1{\langle #1 \rangle}
\newc{\sg}{S}
\newc{\ug}{U}
\newc{\tg}{T}

\allowdisplaybreaks[2]
\begin{document}
\bibliographystyle{OurBibTeX}

\title{\hfill ~\\[-30mm]
                  \textbf{
                  A model of quark and lepton mixing
                                  }        }
\date{}
\author{\\[-5mm]
Stephen F. King\footnote{E-mail: {\tt king@soton.ac.uk}}\\ \\
  \emph{\small School of Physics and Astronomy, University of Southampton,}\\
  \emph{\small Southampton, SO17 1BJ, United Kingdom}\\[4mm]}

\maketitle

\begin{abstract}
\noindent
{We propose a model of quark and lepton mixing based on the tetrahedral $A_4$ family symmetry
with quark-lepton unification via the tetra-colour Pati-Salam gauge group $SU(4)_{PS}$, together with 
$SU(2)_L\times U(1)_R$.
The ``tetra-model'' solves many of the flavour puzzles and remarkably gives ten predictions 
at leading order, including all six PMNS parameters. The Cabibbo angle is approximately 
given by $\theta_C\approx 1/4$, due to the tetra-vacuum alignment $(1,4,2)$,
providing the Cabibbo connection between quark and lepton mixing.
Higher order corrections are responsible for the smaller quark mixing angles and CP violation
and provide corrections to the Cabibbo and lepton mixing angles and phases.
The tetra-model 
involves an $SO(10)$-like pattern of Dirac and heavy right-handed neutrino masses,
with the strong up-type quark mass hierarchy cancelling in the see-saw mechanism, leading to a normal hierarchy of neutrino masses
with an atmospheric angle in the first octant, 
$\theta^l_{23}= 40^\circ \pm1^\circ$, a solar angle
$\theta^l_{12}= 34^\circ \pm1^\circ$,
a reactor angle $\theta^l_{13}= 9.0^\circ \pm 0.5^\circ$, depending on the ratio 
of neutrino masses $m_2/m_3$,
and a Dirac CP violating oscillation phase $\delta^l=260^\circ \pm 5^\circ$.} 
 \end{abstract}
\thispagestyle{empty}
\vfill
\newpage
\setcounter{page}{1}

\section{Introduction}

The discovery of a Higgs boson at the LHC \cite{Aad:2012tfa}
provides convincing evidence for the Standard Model (SM) picture of electroweak
symmetry broken by the vacuum expectation value (VEV) of a doublet of complex scalars.
In the SM, the Higgs doublet is also responsible for quark and charged lepton masses and quark mixing 
via the Yukawa couplings to fermions.
However the SM offers no insight into pattern of such Yukawa couplings, nor into the origin and nature
of neutrino mass. Indeed
it is worth recalling that the flavour sector of the 
SM involves at least twenty undetermined parameters, including 
ten parameters in the quark sector comprising 
the six quark masses, the three quark mixing angles and the phase describing CP violation.
The lepton sector involves at least a further ten physical parameters, comprising the
three charged lepton masses, three neutrino masses, three lepton mixing angles and the phase describing CP violation in the lepton sector. If neutrinos are Majorana, then there will be another two CP violating leptonic phases.
The most recent best fit values of leptonic mixing parameters are \cite{GonzalezGarcia:2012sz}:  
$\theta^l_{12}=34^\circ \pm 0.8^\circ$,
$\theta^l_{23}= 42^\circ \pm 2^\circ$ or $\theta^l_{23}= 50^\circ \pm 2^\circ$,
$\theta^l_{13}= 9^\circ \pm 0.4^\circ$, $\delta^l= 270^\circ \pm 70^\circ$,
where the errors quoted are one sigma ranges.


Following the discovery of neutrino mass and mixing in 1998, 
there has been a major discovery in neutrino physics almost
every year (for an up to date review see e.g.~\cite{King:2013eh}).
For example, in 2012 the reactor angle was measured for the first time,
with the latest central value measured by Daya Bay being $\theta^l_{13}\approx 8.7^{\circ}$ \cite{DayaBay}.
The measurement of the reactor angle 
excluded many neutrino mixing models, and led to new model building strategies based on discrete family symmetries as reviewed in~\cite{King:2013eh}. 
The discoveries in neutrino physics have enriched the flavour puzzle, raising new questions such as 
the smallness of neutrino masses compared to charged fermion masses, the stronger hierarchy of charged fermion masses compared to neutrino masses and
the smallness of the quark mixing angles compared to lepton mixing angles, apart from
the Cabibbo angle $\theta_{C}$ which is of similar size to the reactor angle,
for example $\theta^l_{13}\approx \theta_{C}/\sqrt{2}$ \cite{Giunti:2002ye},
which may be combined with tri-bimaximal (TB) mixing \cite{King:2012vj}.
These new flavour puzzles are in addition to the long standing questions such as 
the similarity of charged lepton masses to down-type quark masses and the stronger hierarchy of up quark masses compared to down quark masses.
The origin of CP violation in both the quark and (so far unmeasured) lepton sectors also remains a mystery.

The see-saw mechanism \cite{Minkowski:1977sc} sheds light on 
the smallness of neutrino masses but can increase the parameter count considerably due to an 
undetermined right-handed neutrino Majorana mass matrix.
In the diagonal right-handed neutrino and charged lepton basis (the so-called flavour basis)
there is an undetermined neutrino Yukawa matrix. 
Without the see-saw mechanism,
the SM involves three charged fermion Yukawa matrices but these
are non-physical and basis dependent quantities.
However, in theories of flavour beyond the SM, the choice of basis may well have physical significance
and, in a certain basis defined by the theory,
the Yukawa matrices may take simple forms, leading to some
predictive power of the model as a result.

Recently we proposed a model of leptons 
\cite{King:2013xba,King:2013iva}
based on the see-saw mechanism in which the number of parameters in the lepton sector was dramatically reduced.
In the flavour basis,
the right-handed neutrino mainly responsible for the atmospheric neutrino mass has couplings to $(\nu_e, \nu_{\mu}, \nu_{\tau})$ proportional to $(0,1,1)$ and the 
right-handed neutrino mainly responsible for the solar neutrino mass has couplings to $(\nu_e, \nu_{\mu}, \nu_{\tau})$ proportional to $(1,4,2)$, with a
relative phase $\eta = \pm 2\pi/5$,
where the couplings and phase originated from vacuum alignment with
$A_4$ and $Z_5$ discrete symmetries. The model involved two right-handed neutrinos as a limiting
case of sequential dominance (SD) \cite{King:1998jw}. The model 
predicted lepton mixing angles 
which agreed very well with the best fit values for a normal neutrino mass hierarchy,
together with predictions for the CP violating phases, whose sign depended on the sign of the
phase $\eta = \pm 2\pi/5$.
The goal of the present paper is to extend the above model of leptons to the quark sector,
in such a way as to preserve the successful predictions in the lepton sector,
thereby providing a complete model of quark and lepton masses and mixing. 

\begin{figure}[t]
\centering
\includegraphics[width=0.46\textwidth]{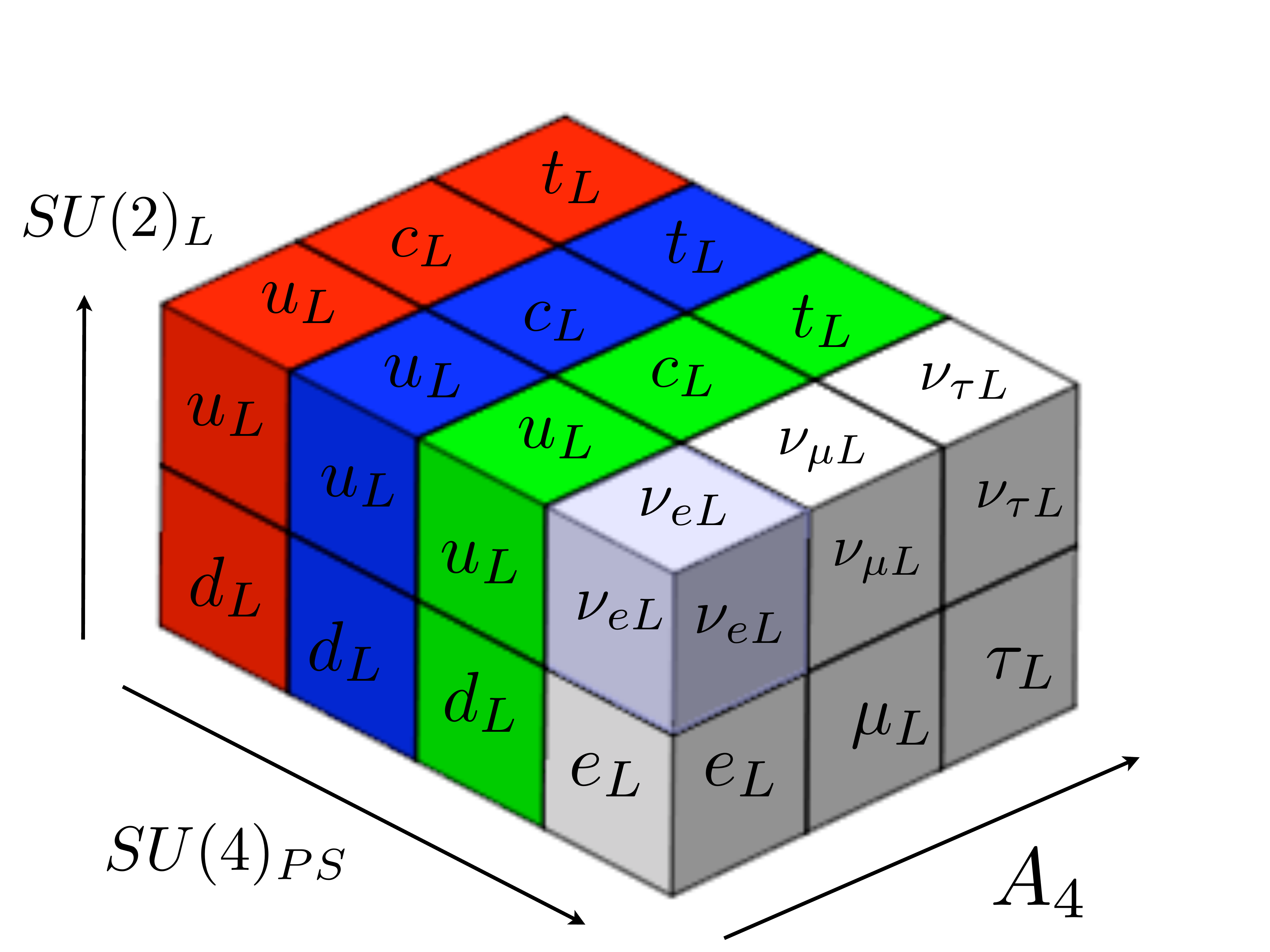}
\includegraphics[width=0.50\textwidth]{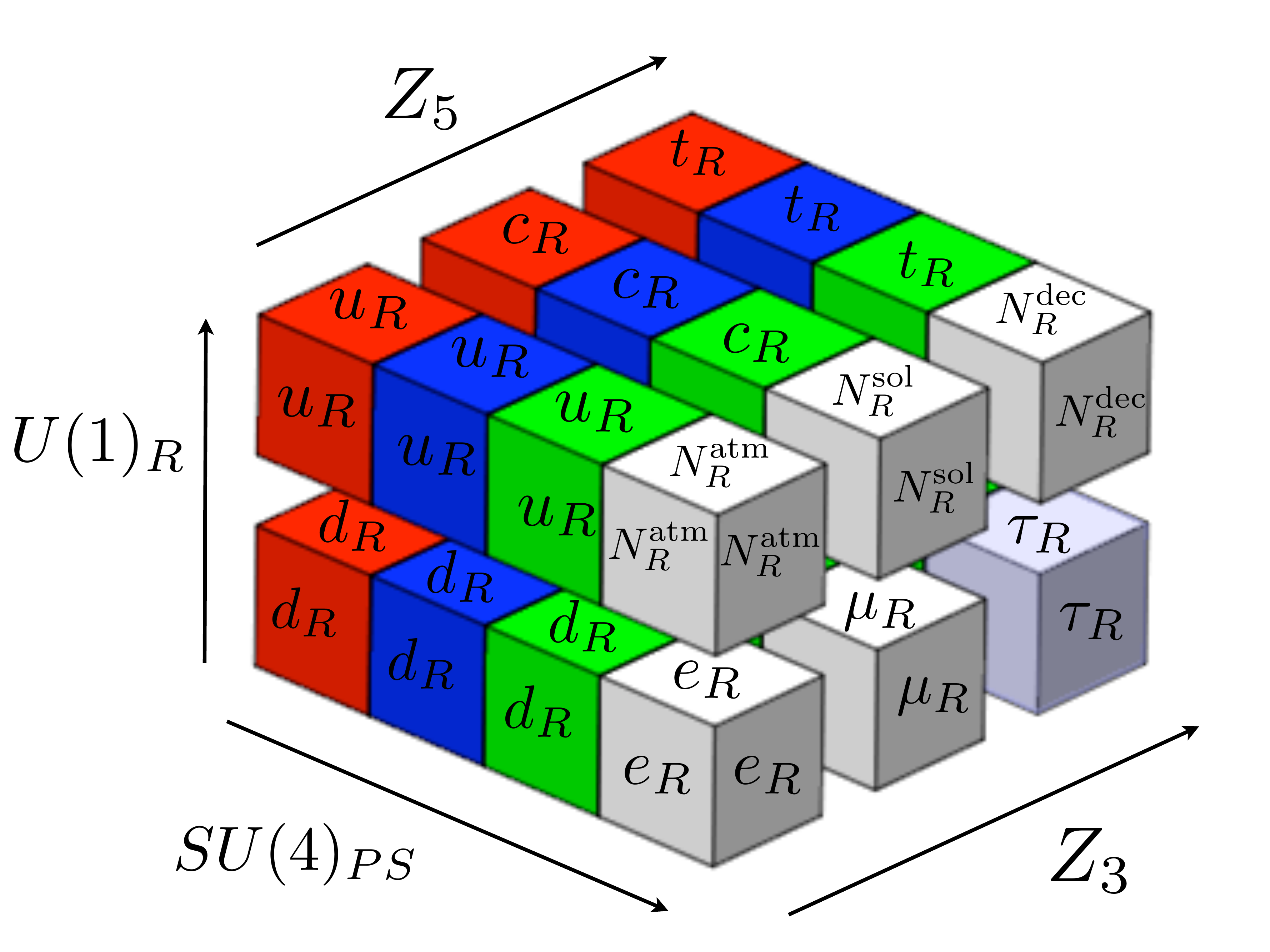}
\vspace*{-4mm}
    \caption{The A4SU421 unification of quarks and leptons in the ``tetra-model''.
    The left diagram depicts quark-lepton-family 
    unification of the 24 left-handed quarks and leptons denoted collectively as ${\cal Q} $ into a single 
   $(3,4,2,0)$ multiplet of A4SU421.
The right diagram shows the 24 right-handed quarks and leptons which form six $A_4$ singlets,
${\cal U}_i$ and ${\cal D}_i$,
distinguished by $Z_5$ and $Z_3$, 
with quarks and leptons unified in each multiplet.
  } \label{421fig}
\vspace*{-2mm}
\end{figure}

In this paper we propose a model of quark and lepton mixing
based on the tetrahedral $A_4$ discrete family symmetry and the tetra-colour 
Pati-Salam (PS) gauge group $SU(4)_{PS}$ \cite{Pati:1974yy},
together with $SU(2)_L\times U(1)_R$, where we refer to this group as A4SU421.
The A4SU421 model with the above tetra-vacuum alignment $(1,4,2)$ will be referred to as the 
``tetra-model'' for brevity. 
The unification of quarks and leptons in terms of A4SU421 
is depicted in 
Fig.~\ref{421fig}.
The model involves $U(1)_R$, rather than $SU(2)_R$,
used in previous models \cite{King:2006np}, in order to allow
diagonal charged lepton and down quark
Yukawa matrices together with off-diagonal neutrino and up quark Yukawa matrices.
Quark mixing then arises completely from the up quark Yukawa matrix,
which is equal to the neutrino Yukawa up to Clebsch-Gordan coefficients.
The diagonal charged lepton and
down quark Yukawa matrices are also equal up to Clebsch-Gordan coefficients
due to the $SU(4)_{PS}$.
The Cabibbo angle is predicted to be $\theta_C\approx 1/4$
due to the tetra-vacuum alignment in the second column $(1,4,2)$, which is common to 
the neutrino and up Yukawa matrices, providing a Cabibbo connection between
quark and lepton mixing. The tetra-model 
predicts an $SO(10)$-like pattern of Dirac and heavy right-handed neutrino masses,
with the strong up-quark mass hierarchy cancelling in the see-saw mechanism, leading to a normal neutrino mass hierarchy.

It is worth discussing how the tetra-model compares to some other recent attempts to explain both quark and lepton
mixing as a result of discrete family symmetry, following the measurement of the reactor angle.
Models can be classified as direct, semi-direct or indirect, depending to what extent a subgroup of the discrete family symmetry can be identified with the Klein symmetry of the neutrino sector~\cite{King:2013eh}. 
In several of these models quarks are included via $SU(5)$
unification, but typically vacuum alignment does not determine the quark mixing angles.
However, in a purely symmetry approach, the direct approach has been extended to the quark sector, where a subgroup of the discrete family symmetry is used to constrain also the quark mixing angles, in analogy with the procedure followed  
for the Klein symmetry in the neutrino sector \cite{Hagedorn:2012pg,Araki:2013rkf,Holthausen:2013vba},
but no realistic model has been proposed. In some such approaches \cite{Araki:2013rkf,Holthausen:2013vba},
the symmetry groups can be quite large, for example $\Delta (6n^2)$ for large values of $n$
\cite{King:2013vna}. 

Here we follow the indirect approach where small discrete family symmetries 
such as $A_4$ are used to 
facilitate interesting vacuum alignments, rather than as the direct origin of the Klein
symmetry. Including the SD mechanism \cite{King:1998jw} and vacuum alignment,
various forms of constrained sequential dominance (CSD) have been considered based on the 
atmospheric neutrino alignment $(0,1,1)$ but with different forms of 
solar neutrino alignment: original CSD \cite{King:2005bj} involved a solar alignment
$(1,1,-1)$ yielding tri-bimaximal (TB) mixing; CSD2 \cite{Antusch:2011ic} involved a solar alignment $(1,2,0)$ and hence a small reactor angle;
CSD3 \cite{King:2013iva}
involved solar alignment $(1,3,1)$ with an acceptable reactor angle but maximal atmospheric mixing;
CSD4 \cite{King:2013xba} with the tetra-alignment 
$(1,4,2)$ adopted here predicts best fit lepton angles with a normal hierarchy. By unifying leptons with quarks, we 
show here for the first time that CSD4 can also successfully predict the Cabibbo angle.

The layout of the remainder of the paper is as follows. In section~\ref{model} we introduce the tetra-model,
and show how the vacuum alignments imply the pattern of Yukawa matrices described above,
in the down, up and Majorana sectors. 
In section~\ref{leading} we collect together all the Yukawa matrices
and discuss the leading order predictions of the model, first qualitatively, then giving the quantitative
predictions in the lepton sector in the presence of the third right-handed neutrino leading to a non-zero
lightest neutrino mass. 
In section~\ref{higher} we discuss the higher order corrections to the model,
responsible for the small quark mixing angles and CP violation,
first studying the operators, then the effect of these operators on the 
Yukawa matrices and hence on the predictions for all
quark and lepton masses and mixing angles. 
Section~\ref{conclusions} concludes the paper.

\section{The tetra-model}
\label{model}

\begin{table}
{	\centering
$$
\begin{array}{||c||ccccccccccc||cccc||}
\hline \hline
& {\cal Q} & {\cal U}^c_i & {\cal D}^c_i & \phi_{{\cal U}^c_i} & \phi_{{\cal D}^c_i} & {\cal H}_{\cal U} & \overline{\cal H}_{\cal U} & h_u & h_d & h_{\cal D} & h_{\cal U}
& \Sigma_{15}& \Sigma'_{15}&\overline{X}_{{\cal Q}_i}&{X}_{{\cal Q}_i}
\\   \hline  \hline 
 A_4 & 3 & 1 & 1 & 3 & 3 & 1 & 1 & 1 & 1 & 1 & 1&1&1&1&1\\[2mm] 
  SU(4)_{PS} & 4 & \overline{4} & \overline{4} & 1 & 1 & \overline{4} & 4 & 1 & 1 & 15 & 15 &15&15& \overline{4}&4\\[2mm]
   SU(2)_L & 2 & 1 & 1 & 1 & 1 & 1 & 1 & 2 & 2 & 2 & 2 &1&1&2&2\\[2mm]
    U(1)_R & 0 & -\frac{1}{2} & \frac{1}{2} & 0 & 0 & -\frac{1}{2}&  \frac{1}{2} &  \frac{1}{2} &  -\frac{1}{2}
    & -\frac{1}{2}&  \frac{1}{2} &0&0&0&0\\[2mm]
 \hline \hline
\end{array}
$$
}
\caption{\label{A4PS}
Fields and their transformation properties under $A_4$  and Pati-Salam symmetries.
Fields not shown in this table (for example $\Sigma$) are singlets under $A_4$  and Pati-Salam symmetries.
}
\end{table}

\subsection{Overview}
The model is based on tetrahedral $A_4$ family symmetry combined with 
the tetra-colour Pati-Salam gauge group $SU(4)_{PS}$ together with $SU(2)_L \times U(1)_R$,
\beq
A_4\times SU(4)_{PS} \times SU(2)_L \times U(1)_R,
\eeq
where we refer to this group as A4SU421.
Formally $U(1)_R$ may be identified as the diagonal subgroup of the Pati-Salam right-handed gauge 
group $SU(2)_R$ with $R=T_{3R}$, the third generator of $SU(2)_R$. 
However we only assume a $U(1)_R$ gauge group since we require diagonal down and charged lepton
Yukawa matrices together with off-diagonal up and neutrino Yukawa matrices, and this is very difficult to 
achieve if the full $SU(2)_R$ is respected. For the same reason it is not possible to embed the model into 
$SO(10)$. An additional reason why the $SO(10)$ embedding is not possible is that the left-handed and right-handed quarks and leptons transform differently under $A_4$, as discussed below.

The left-handed quarks and leptons are unified into the single multiplet ${\cal Q}$ 
while the (CP conjugated) right-handed fields ${\cal U}^c_i$ and ${\cal D}^c_i$
are $A_4$ singlets, transforming under A4SU421 as,
\beq
{\cal Q}=(3,4,2,0), \ \ {\cal U}^c_i = (1,\overline{4},1,-1/2), \ \  {\cal D}^c_i = (1,\overline{4},1,1/2).
\eeq
The unification of quarks and leptons has already been depicted in 
Fig.~\ref{421fig}.
The full list of fields which transform under the $A_4$ and/or the Pati-Salam group
are shown in Table~\ref{A4PS}.
Clearly above tetra-model cannot be embedded into $A_4\times SO(10)$ since different components 
of the 16-dimensional representation of $SO(10)$ transform differently under $A_4$.

The partial Pati-Salam gauge group is broken to the SM,
\beq
SU(4)_{PS}\times U(1)_R\rightarrow SU(3)_C\times U(1)_{B-L}\times U(1)_R \rightarrow SU(3)_C\times U(1)_Y,
\eeq
by PS Higgs,
${\cal H}_{\cal U}=(H_{U^c},H_{N^c})$  and $\overline{\cal H}_{\cal U}=(\overline{H}_{U^c},\overline{H}_{N^c})$, 
which acquire VEVs in the ``right-handed neutrino'' directions $\vev{H_{N^c}}=\vev{\overline{H}_{N^c}}$.
If the breaking occurs at high scales, close to $2\times 10^{16}$ GeV, then supersymmetric gauge coupling unification of the SM gauge couplings is maintained.
The preserved hypercharge generator is given by,
\beq
Y=\frac{B-L}{2}+R.
\eeq

The choice of gauge group SU421 has been recently considered in \cite{Perez:2013osa},
although without any discrete family symmetry such as $A_4$ considered here.
It is worth pointing out that in the tetra model none of the Higgs fields carry any $A_4$ charges,
while none of the flavons carry any SU421 charges. This means that, in the absence 
of any other flavour symmetries, the 
flavon potential relevant for $A_4$ breaking is independent of the SU421 breaking potential,
where the latter was considered in \cite{Perez:2013osa} for a non-supersymmetric model. 
The potential for the minimal 
supersymmetric SU422 potential has been considered in \cite{King:1997ia}.

Below the PS scale, $h_{\cal D} \sim (15,2,-1/2)$ will yield 
a Higgs doublet with the same quantum numbers as $h_d \sim (1,2,-1/2)$.
When the resulting mass matrix of Higgs doublets is diagonalised, there will be a 
single low energy down-type Higgs doublet consisting of a mixture of 
the Higgs doublet contained in $h_{\cal D}\sim (15,2,-1/2)$ and $h_{d}\sim (1,2,-1/2)$.
This is of course a well known effect \cite{Georgi:1979df}. 
A similar mass mixing may also arise between the Higgs doublet in 
$h_{\cal U}\sim (15,2,1/2)$ and $h_{u}\sim (1,2,1/2)$ leading to a single low energy
up-type Higgs doublet. We therefore expect two low energy electroweak Higgs doublets,
one up-type and one down-type, as in the MSSM.

However the use of such minimal Higgs potentials has been called into question in theories 
where some of the Higgs fields transform under both GUT and flavour symmetries \cite{Toorop:2010yh}.
In the present model it will turn out that $h_{\cal D} \sim (15,2,-1/2)$ and $h_{\cal U} \sim (15,2,1/2)$
will transform under $Z_3^{\cal D}$ and $Z_5^{\cal U}$ flavour symmetries, and these charges
will require the standard Higgs potentials to be modified. In the present model it will turn out that 
the combination of Higgs fields $h_{\cal D}\Sigma_{15}$ has exactly the same quantum numbers under all
symmetries (including flavour symmetries) as $h_d$. Similarly the combination of Higgs fields $h_{\cal U}\Sigma'_{15}$ has exactly the same quantum numbers under all symmetries as $h_u$. Therefore, the standard Higgs potentials may be used together with extra non-renormalisable terms obtained by replacing $h_d\rightarrow h_{\cal D}\Sigma_{15}$ and $h_u\rightarrow h_{\cal U}\Sigma'_{15}$.
When $\Sigma'_{15}$ and $\Sigma_{15}$ develop vacuum expectation values, 
the extra terms yield the desired Higgs mixing as in the standard mechanisms without flavour symmetry.

The $A_4$ is broken by the VEVs of six triplet flavons $\phi_{{\cal U}^c_i}$ and $\phi_{{\cal D}^c_i}$,
which couple in a one-one correspondence with ${\cal U}^c_i$ and ${\cal D}^c_i$.
The remaining fields are messengers entering 
the diagrams in  Figure~\ref{mess} as discussed later.

\subsection{CSD4 Vacuum Alignments}
The structure of the Yukawa matrices depends on the so-called CSD4 vacuum alignments
which were first derived in \cite{King:2013xba},
\be
\langle \phi_{{\cal U}^c_1} \rangle =
\frac{v_{{\cal U}^c_1}}{\sqrt{2}}  \begin{pmatrix}0 \\ 1 \\ 1\end{pmatrix}  \ , \qquad
\langle \phi_{{\cal U}^c_2} \rangle =
\frac{v_{{\cal U}^c_2}}{\sqrt{21}} \begin{pmatrix}1 \\ 4 \\ 2\end{pmatrix} \ , \qquad
\langle \phi_{{\cal U}^c_3} \rangle = v_{{\cal U}^c_3}
\begin{pmatrix} 0 \\ 0 \\1 \end{pmatrix} 
 \ ,\label{phiu}
\ee
and
\be
\langle \phi_{{\cal D}^c_1} \rangle =v_{{\cal D}^c_1}
\begin{pmatrix} 1\\0\\0 \end{pmatrix}  \ , \qquad
\langle \phi_{{\cal D}^c_2} \rangle =v_{{\cal D}^c_2}
\begin{pmatrix} 0\\1\\0 \end{pmatrix} \ , \qquad
\langle \phi_{{\cal D}^c_3} \rangle = v_{{\cal D}^c_3}
\begin{pmatrix} 0\\0\\1 \end{pmatrix} .
 \ \label{phid}
\ee
The mechanism for the vacuum alignment, especially the tetra-alignment $(1,4,2)$,
relies mainly on othogonality of flavons as
discussed in \cite{King:2013xba}. It is noteworthy that 
we impose a CP symmetry which is spontaneously broken by VEVs of the flavons.
Due to the $Z_5$ symmetries, the $ \phi_{{\cal U}^c_i}$ flavons can only acquire a discrete choice of overall phase
corresponding to some multiple of $2\pi/5$.
Similarly, due to the $Z_3$ symmetries, the $ \phi_{{\cal D}^c_i}$ flavons can only acquire a discrete choice of overall phase corresponding to some multiple of $2\pi/3$.
As in \cite{King:2013xba}, we will select all the phases of the triplet flavons to be all zero, with CP violation originating
from the phases of the singlet flavons $\xi_i$ as discussed later.

At leading order, the CSD4 vacuum alignment of the flavons, together with 
operators of the form $(\phi_{{\cal U}^c_i}.{\cal Q}){\cal U}^c_i$ and 
$(\phi_{{\cal D}^c_i}.{\cal Q}){\cal D}^c_i$,
imply that the Yukawa matrices (in LR convention) are constructed from the column vectors above.

The up and neutrino Yukawa matrices 
are obtained from $(\phi_{{\cal U}^c_i}.{\cal Q}){\cal U}^c_i$ 
by sticking together the 
three column vectors in Eq.\ref{phiu},
\be
  Y^{\nu} \sim Y^u \sim  \begin{pmatrix}  0 & b & 0  \\ 
a & 4b & 0\\  a & 2b & c\end{pmatrix},
\ee
where each column is multiplied by a different constant of proportionality.
The Yukawa matrices are not expected to be exactly equal due to Clebsch-Gordan coefficients,
as discussed later.

The down and charged lepton Yukawa matrices are similarly obtained from
$(\phi_{{\cal D}^c_i}.{\cal Q}){\cal D}^c_i$
by amalgamating the 
three column vectors
in Eq.\ref{phid} and are hence diagonal,
\be
 Y^d \sim Y^e \sim
  \begin{pmatrix}  y_d & 0  & 0  \\ 0 & y_s & 0\\ 0  & 0 & y_b\end{pmatrix}.
\label{phid1}
\ee
As mentioned above, the Yukawa matrices are not expected to be exactly equal due to Clebsch-Gordan coefficients,
as discussed later.

The quark-lepton unification implies that 
the second column $(1,4,2)^T$ of the neutrino Yukawa matrix is equal to that of the up quark Yukawa
matrix and hence predicts a Cabibbo angle approximately equal to 1/4.
The third column (approximately decoupled from the see-saw mechanism) 
is proportional to $(0,0,1)^T$ at leading order
giving the top quark Yukawa coupling. 
Higher order corrections modify the leading order predictions
and are responsible for 
the other quark mixing angles and CP violation.

As discussed in the following subsections, 
the model employs other auxiliary $Z_5$ and $Z_3$ symmetries in order to ensure the one-one correspondence
of the couplings of the flavons $\phi_{{\cal U}^c_i}$ and $\phi_{{\cal D}^c_i}$ with ${\cal U}^c_i$ and ${\cal D}^c_i$
in the Yukawa operators.
These symmetries also predict Clebsch-Gordan relations between the down quark and charged lepton 
masses, as well as the up quark mass hierarchy, with the charges cancelling in the see-saw mechanism, leading to a mild normal neutrino mass hierarchy.
However right-handed neutrino masses are predicted to be very hierarchical, being proportional to the squares of up-type quark masses, which is another consequence of quark-lepton unification.

\subsection{The down sector}
\begin{figure}
\centering
\includegraphics[width=0.3\textwidth]{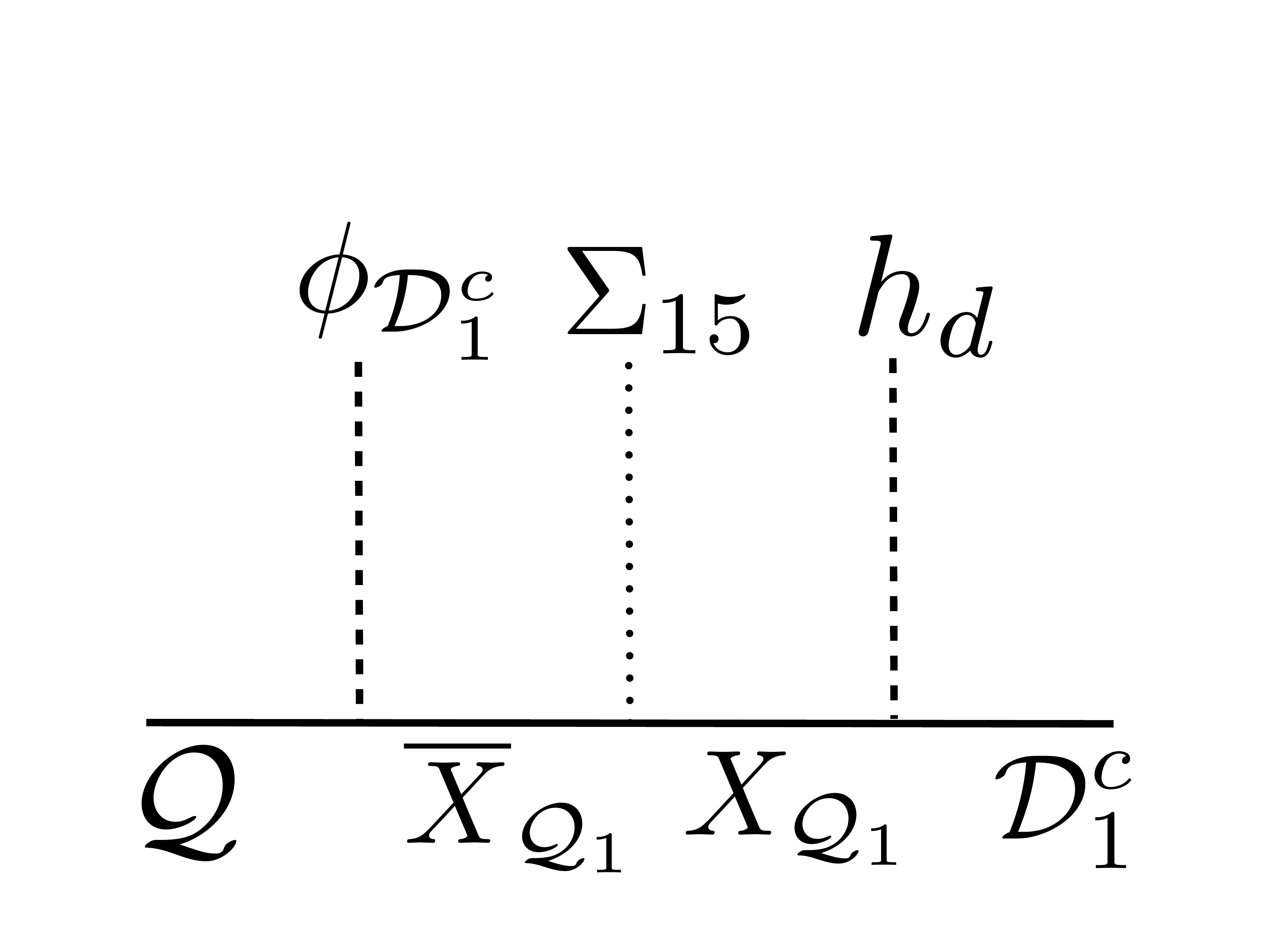}
\includegraphics[width=0.3\textwidth]{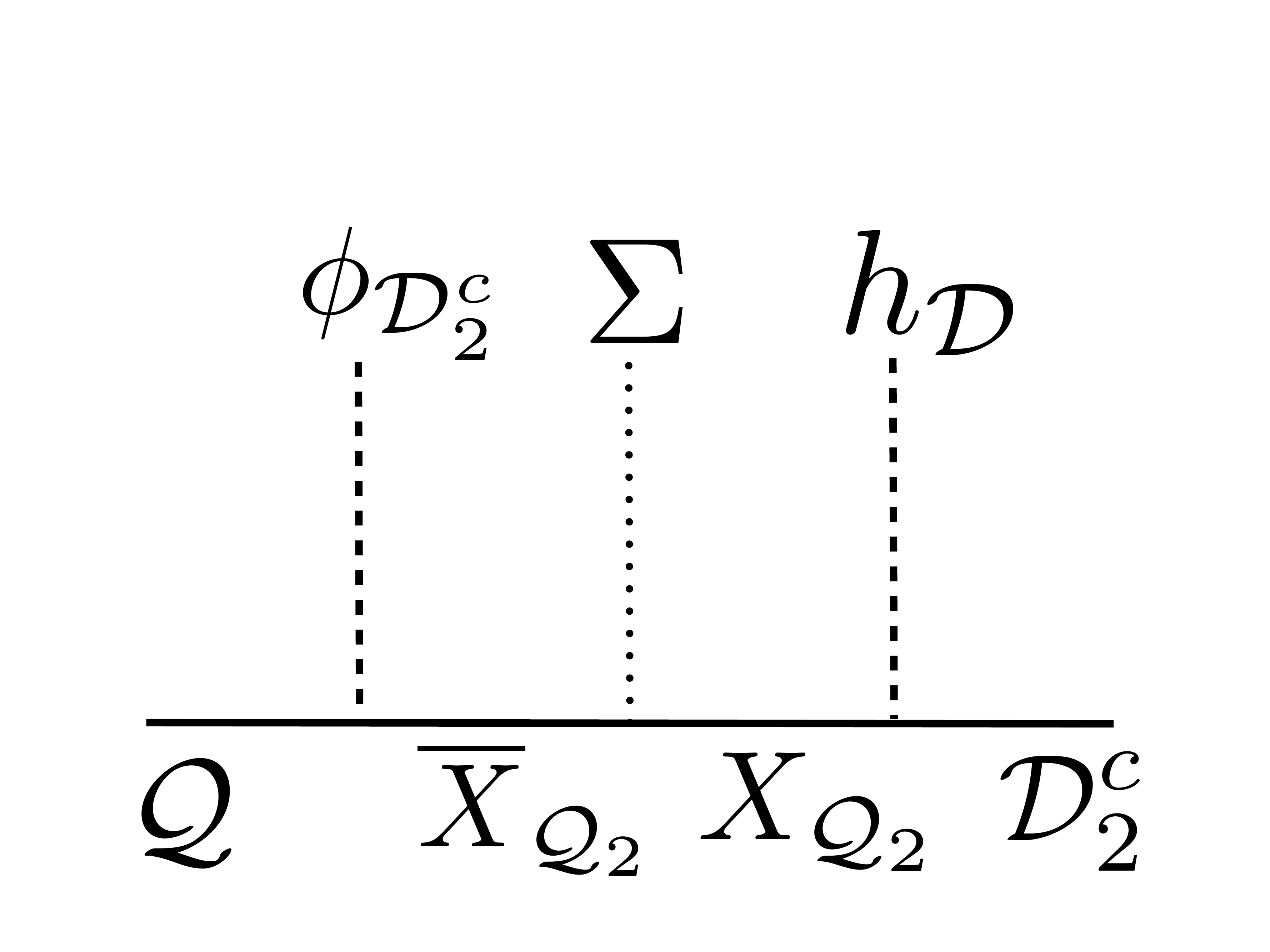}
\includegraphics[width=0.3\textwidth]{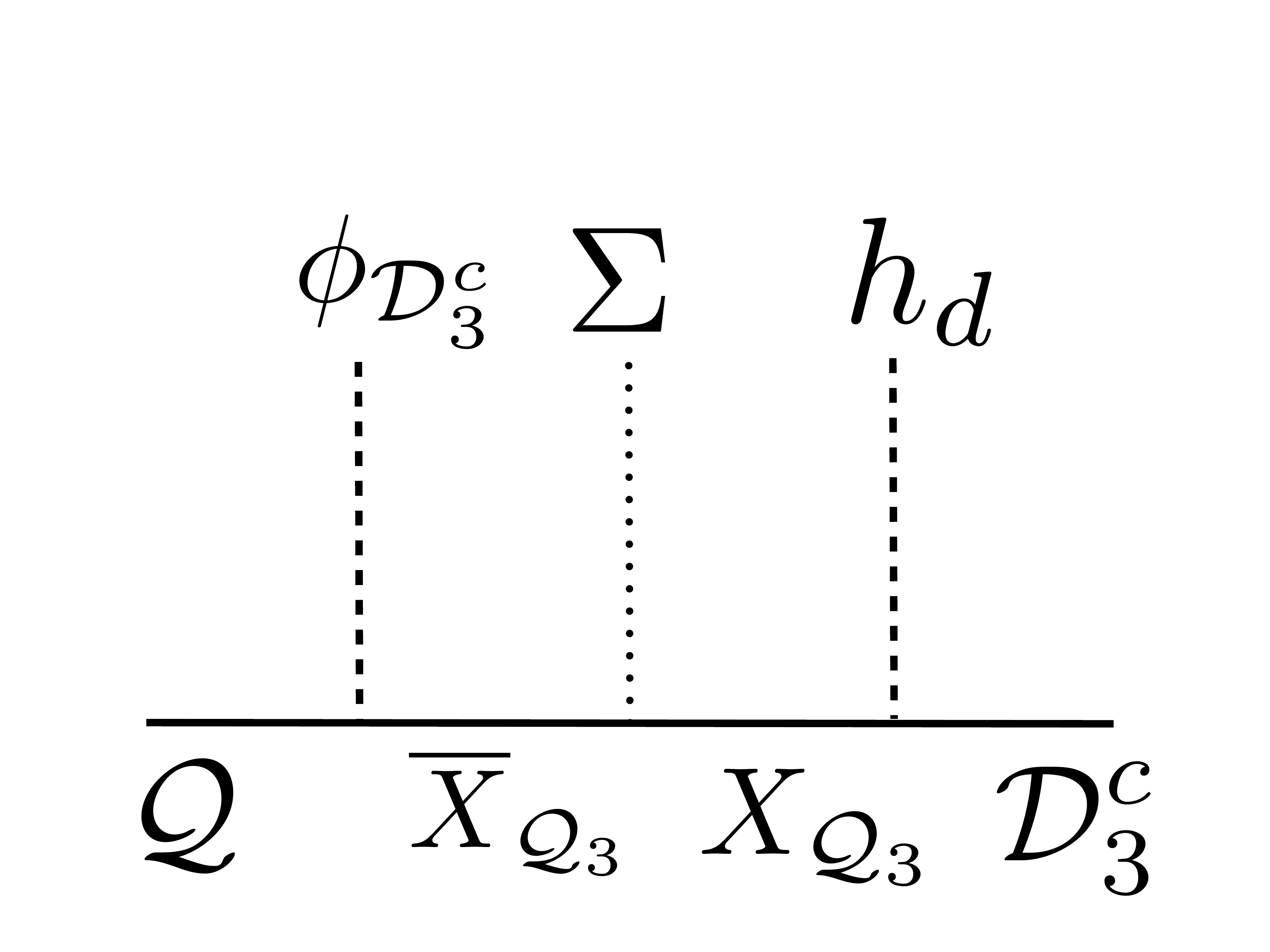}
\vspace*{-4mm}
    \caption{These diagrams show the messenger sector responsible for the 
    effective operators in Eq.\ref{Yukd} responsible for the
    charged lepton and down quark masses.     } \label{mess}
\vspace*{-2mm}
\end{figure}

We first consider the down sector, 
where the postulated $Z_3$ symmetries and charges are shown in Table~\ref{tabD}.
The $Z_3^{{\cal D}^c_i}$ are used to make the Yukawa operators diagonal (i.e. to stick
a particular flavon $\phi_{{\cal D}^c_i}$ to a particular matter field ${\cal D}^c_i$).
The $Z_3^{\cal D}$ is used to control the down messenger sector of the model
leading to the diagrams in Figure~\ref{mess}.

The $Z_3$ allowed effective operators, which result from the diagrams in Figure~\ref{mess}
below the scales $\vev{\Sigma}$ 
and $\vev{\Sigma_{\cal D}}$ (which are assumed to be
higher than the $A_4$ breaking scale) are
\beq
\frac{y^{\cal D}_{1}}{\vev{\Sigma_{15}}}h_d(\phi_{{\cal D}^c_1}.{\cal Q}){\cal D}^c_1
+\frac{y^{\cal D}_{2}}{\vev{\Sigma}}h_{\cal D}(\phi_{{\cal D}^c_2}.{\cal Q}){\cal D}^c_2
+\frac{y^{\cal D}_{3}}{\vev{\Sigma}}h_d(\phi_{{\cal D}^c_3}.{\cal Q}){\cal D}^c_3,
\label{Yukd}
\eeq
where we have introduced a new PS Higgs in the adjoint of $SU(4)_{PS}$,
$h_{\cal D}\sim (15,2,-1/2)$, which couples to ${\cal D}^c_2$, leading a Clebsch factor of 3 
between charged lepton and down-type quark masses for the second family \cite{Georgi:1979df}. 
The messenger mass for the first family arises from
the coupling to an adjoint of $SU(4)_{PS}$, $\Sigma_{\cal D}\sim (15,1,0)$,
giving a Clebsch factor of 3 in the denominator \cite{Antusch:2013rxa}.
The messenger mass for the second and third families arises from the 
PS singlet $\Sigma \sim (1,1,0)$ so no inverse Clebsch factors arise in these cases.

The resulting Yukawa matrices are diagonal and given by,
\begin{equation} \label{Yed}
Y^d =   \begin{pmatrix}  y_d & 0  & 0  \\ 0 & y_s & 0\\ 0  & 0 & y_b\end{pmatrix},
\ \ \ \ 
Y^e =  \begin{pmatrix}  y_d/3 & 0  & 0  \\ 0 & 3y_s & 0\\ 0  & 0 & y_b\end{pmatrix},
\end{equation}
where
\beq
y_d=\frac{y^{\cal D}_{1}v_{{\cal D}^c_1}}{\vev{\Sigma_{15}}} , \ \ 
y_s=  \frac{y^{\cal D}_{2}\epsilon_{\cal D}v_{{\cal D}^c_2}}{\vev{\Sigma}} , \ \ 
y_b= \frac{y^{\cal D}_{3}v_{{\cal D}^c_3}}{\vev{\Sigma}},
\eeq
where we have included a small mixing parameter $\epsilon_{\cal D}$ 
\footnote{The smallness of the parameter $\epsilon_{\cal D}$ may be naturally explained since this 
mixing arises from non-renormalisable operators as discussed earlier.}
associated with the
high energy mixing of the Higgs doublet arising from $h_{\cal D}\sim (15,2,-1/2)$ 
with that
in $h_d\sim (1,2,-1/2)$, which may account for the smallness of the second family masses.
The down-type quark and charged lepton masses are then given by,
\beq
m_e=\frac{m_d}{3}, \ \ m_{\mu}=3m_s, \ \ m_{\tau}=m_b .
\eeq
These are the well-known Georgi-Jarlskog (GJ) relations \cite{Georgi:1979df}, although here they 
arise from a new mechanism, namely due to non-singlet fields which appear
in the denominator of effective operators and split the messenger masses \cite{Antusch:2013rxa}.
The viablity of the GJ relations is discussed in 
\cite{Antusch:2013jca}.
The smallness of the down quark mass compared to the bottom quark mass is ascribed to the different couplings and VEVs involved in the ratio $y_d/y_b$, for example
by assuming a small ratio $\vev{\Sigma}/\vev{\Sigma_{15}}\ll 1$.

\begin{table}
{	\centering
$$
\begin{array}{||c||ccccccc||cccccccc||}
\hline \hline
&h_{\cal D} 
&{\cal D}^c_1&{\cal D}^c_2& {\cal D}^c_3 &\phi_{{\cal D}^c_1}&\phi_{{\cal D}^c_2}&\phi_{{\cal D}^c_3} 
& \Sigma & \Sigma_{15}&\overline{X}_{{\cal Q}_1}&{X}_{{\cal Q}_1}&\overline{X}_{{\cal Q}_2}&{X}_{{\cal Q}_2}
&\overline{X}_{{\cal Q}_3}&{X}_{{\cal Q}_3}
\\  \hline
\hline
Z_3^{\cal D} & \omega^2 &  \omega  & \omega^2  & 1 &  1 &  \omega & \omega^2 & \omega^2 & \omega
& 1 & \omega^2 &  \omega^2 & \omega^2 & \omega & 1
\\[2mm] 
Z_3^{{\cal D}^c_1} & 1 &  \omega^2  & 1  & 1 &  \omega & 1& 1 & 1 & 1 & \omega^2 & \omega & 1 & 1&1&1\\[2mm] 
Z_3^{{\cal D}^c_2} & 1 &  1   &   \omega^2  & 1 & 1 &  \omega & 1 & 1 & 1& 1 & 1& \omega^2 & \omega &1&1\\[2mm] 
Z_3^{{\cal D}^c_3} & 1  & 1  & 1 &   \omega^2 & 1 & 1 &  \omega & 1 & 1 & 1 & 1& 1 & 1&\omega^2&\omega \\[2mm] \hline \hline
\end{array}
$$
}
\caption{\label{tabD}Fields which transform under the $Z_3$ symmetries which control the down sector
(where $\omega = e^{i2\pi /3}$).
Fields not shown in this table (for example $h_d$) are singlets under all $Z_3$ symmetries.}
\end{table}
%
\begin{figure}
\centering
\includegraphics[width=0.4\textwidth]{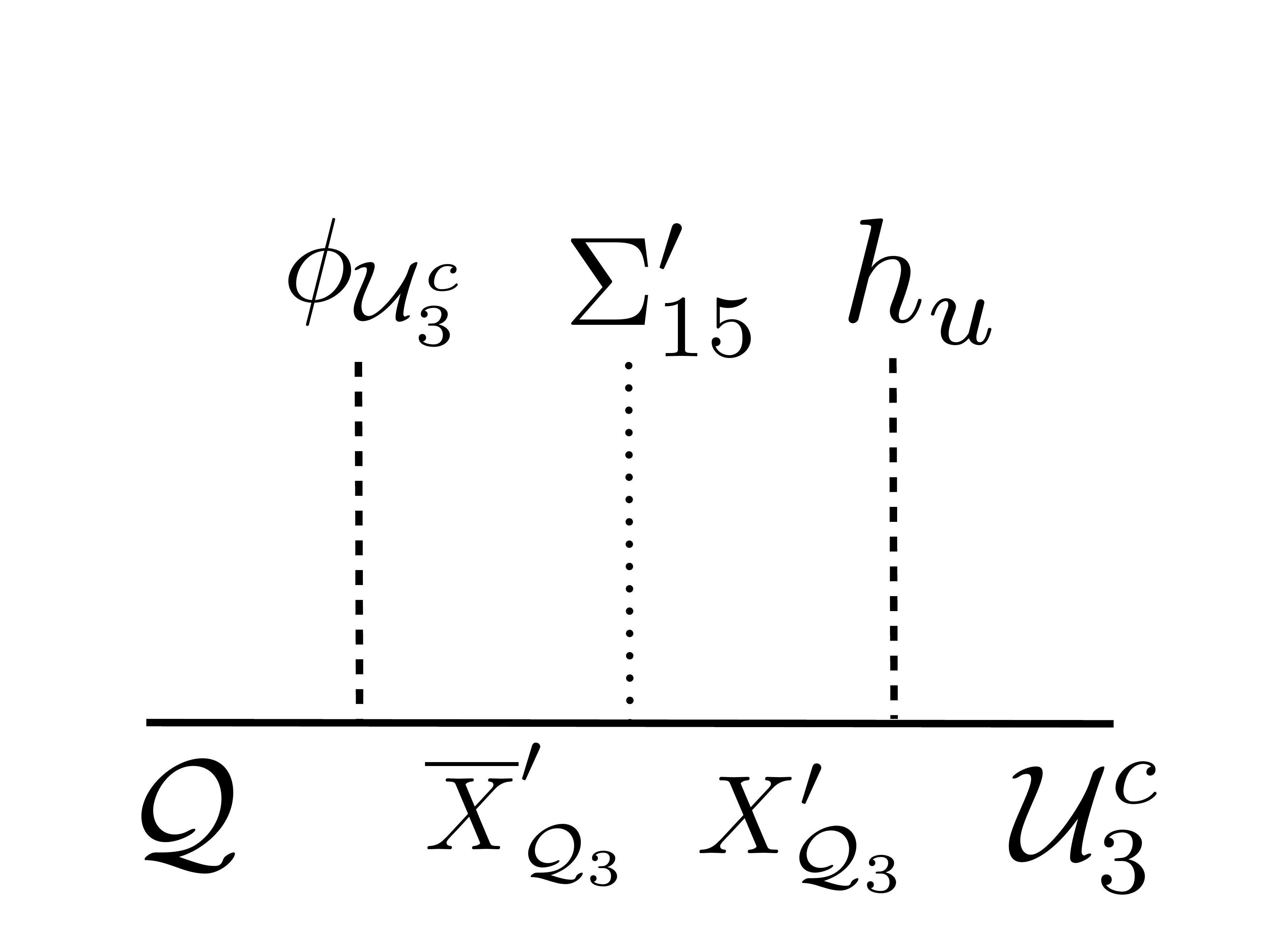}
\vspace*{-4mm}
    \caption{This diagram shows the messenger sector responsible for the 
    effective operator in Eq.\ref{Yuku} responsible for the
     top quark mass and third family Dirac neutrino mass.     } \label{mess2}
\vspace*{-2mm}
\end{figure}

%
\begin{table}
{	\centering
$$
\begin{array}{||c||ccccccccccc||cccc||}
\hline \hline
& h_{\cal U} &\theta_{\cal U} &{\cal U}^c_1&{\cal U}^c_2& {\cal U}^c_3 
&\phi_{{\cal U}^c_1}&\phi_{{\cal U}^c_2}&\phi_{{\cal U}^c_3}
 &\xi_1&\xi_2&\xi_3&\Sigma' & \Sigma'_{15}&\overline{X}'_{{\cal Q}_3}&{X}'_{{\cal Q}_3} \\   \hline  \hline 
 Z_5^{{\cal U}}  &  \rho^4 &  1  & \rho^2 & \rho^2  &\rho &  1 & 1 & 1& \rho &\rho & \rho^3 & \rho^2 & \rho &1 & \rho^4 \\[2mm] 
Z_5^{\theta_{\cal U}} &  1 &  \rho & \rho^3 & \rho^4  &1 &  1 & 1 & 1& 1 &1 & 1 &1 &1 & 1 & 1\\[2mm] 
Z_5^{{\cal U}^c_1}& 1 &   1 & \rho^2  &1 &   1 &\rho^3 & 1 & 1&\rho &1& 1& 1 & 1 & 1 & 1\\[2mm] 
Z_5^{{\cal U}^c_2} & 1 & 1 &1  & \rho^2 & 1 & 1 &\rho^3 & 1&1 &\rho &1 & 1 & 1 & 1 & 1\\[2mm]
Z_5^{{\cal U}^c_3} & 1 & 1  &  1 &1& \rho^2 & 1&1& \rho^3  &1 &1&\rho& 1 & 1 & \rho^2  & \rho^3  \\%
 \hline \hline
\end{array}
$$
}
\caption{\label{tabU}
Fields which transform under the $Z_5$ symmetries which control the up sector
(where $\rho = e^{i2\pi /5}$).
Fields not shown in this table (for example ${\cal Q}$) are singlets under all $Z_5$ symmetries.}
\end{table}
%


\subsection{The up sector}
We now turn to the up sector where the $Z_5$ symmetries are shown in Table~\ref{tabU}.
The $Z_5^{{\cal U}^c_i}$ are used to make the Yukawa operators diagonal (i.e. to stick
a particular flavon $\phi_{{\cal U}^c_i}$ to a particular matter field ${\cal U}^c_i$).
The $Z_5^{\theta_{\cal U}}$ is used to generate the pronounced mass hierarchy in the up sector,
via powers of the flavon field $\theta_{\cal U}$, which is a singlet of both $A_4$ and the Pati-Salam group.
Since the messenger sector in the up sector is more cumbersome than that in the down sector, involving the additional flavons $\theta_{\cal U}$, we only show the operator responsible for the top quark 
and third family neutrino Yukawa coupling in 
Fig.\ref{mess2}.
We highlight an important feature of the messenger sector, namely the presence of 
a symmetry $Z_5^{\cal U}$ which ensures that 
the third family involves a messenger mass term arising from
$\vev{\Sigma'_{15}}$, while the first two families involve messenger masses proportional to $\vev{\Sigma'}$.
This implies that the Dirac mass of the third family neutrino is 1/3 that of the top quark, leading to a normal
neutrino mass hierarchy, as we discuss later.

The leading order $Z_5$ allowed effective operators are,
\footnote{With an alternative choice of charges some of these operators may involve the
Higgs $h_{\cal U}\sim (15,2,1/2)$ 
leading to Clebsch Gordan coefficients analogous to those appearing in the down sector.
For example if ${\cal U}^c_2$ is assigned a $Z_5^{\cal U}$ charge of 
$\rho^3$, with all other charges unchanged, then 
the second operator in Eq.\ref{Yuku} 
will involve $h_{\cal U}$ instead of $h_u$, leading to a Clebsch Gordan coefficient
of 3 multiplying the second column of the neutrino Yukawa matrix $Y^{\nu}$
in Eq.\ref{eq:Ynuu}, and hence $m^D_{\nu2}=3m_c$. }
\beq
\frac{y^{\cal U}_{1}}{\vev{\Sigma'}}\frac{\theta_{\cal U}^2}{\Lambda^2}h_u(\phi_{{\cal U}^c_1}.{\cal Q}){\cal U}^c_1
+\frac{y^{\cal U}_{2}}{\vev{\Sigma'}}\frac{\theta_{\cal U}}{\Lambda}h_u(\phi_{{\cal U}^c_2}.{\cal Q}){\cal U}^c_2
+\frac{y^{\cal U}_{3}}{\vev{\Sigma'_{15}}}h_u(\phi_{{\cal U}^c_3}.{\cal Q}){\cal U}^c_3.
\label{Yuku}
\eeq
For example, below the PS and $Z_5^{\theta_{\cal U}}$ and $Z_5^{\cal U}$
breaking scales, the operators relevant for the neutrino Yukawa matrix $Y^{\nu}$, emerging from 
Eq.~\ref{Yuku}, can be written in a more suggestive notation as,
\begin{equation}
\frac{y^{\cal U}_{1}}{\vev{\Sigma'}}\epsilon^2h_u (\phi_{\rm atm} \cdot L) N^c_{\rm atm}  
+\frac{y^{\cal U}_{2}}{\vev{\Sigma'}}\epsilon h_u (\phi_{\rm sol} \cdot L) N^c_{\rm sol}
+ \frac{y^{\cal U}_{3}}{\vev{\Sigma'_{15}}}h_u (\phi_{\rm dec} \cdot L) N^c_{\rm dec}  ,
\label{ynuops}
\end{equation}
where we have written $\phi_{\rm atm}  \equiv \phi_{{\cal U}^c_1}$, 
 $\phi_{\rm sol}  \equiv \phi_{{\cal U}^c_2}$, 
  $\phi_{\rm dec}  \equiv \phi_{{\cal U}^c_3}$
  and $\epsilon =  \frac{\langle \theta_{\cal U}\rangle }{\Lambda}$. 
  Since these triplet flavons acquire real VEVs these operators will result in a real neutrino Yukawa matrix.

With the vacuum alignments in Eq.~\ref{phiu},
the operators in Eq.~\ref{Yuku} therefore result in the Yukawa matrices,
\begin{equation} \label{eq:Ynuu}
 Y^u =  \begin{pmatrix}  0 & b\epsilon  & 0  \\ 
a\epsilon^2 & 4b\epsilon & 0\\  a\epsilon^2 & 2b\epsilon & c\end{pmatrix},\ \ 
Y^{\nu} = \begin{pmatrix}  0 & b\epsilon  & 0  \\ 
a\epsilon^2 & 4b\epsilon & 0\\  a\epsilon^2 & 2b\epsilon & c/3\end{pmatrix},
\end{equation}
where 
\beq
a=\frac{y^{\cal U}_{1}v_{{\cal U}^c_1}}{\sqrt{2}\vev{\Sigma'} } , \ \ 
b=  \frac{y^{\cal U}_{2} v_{{\cal U}^c_2}}{\sqrt{21}\vev{\Sigma'} } , \ \ 
c= \frac{y^{\cal U}_{3}v_{{\cal U}^c_3}}{  \vev{\Sigma'_{15}}   }
\label{abc}
\eeq
Note that the large top mass implies,
\beq
c= \frac{y^{\cal U}_{3}v_{{\cal U}^c_3}}{  \vev{\Sigma'_{15}}   } \sim 1.
\eeq
This implies that $\langle \phi_{{\cal U}^c_3} \rangle \sim  \vev{\Sigma'_{15}}$
and hence the messenger mass in Fig.\ref{mess2} is of the same
order as the flavon VEV.

The hierarchy of up-type quark masses is controlled by the small parameter $\epsilon$,
and assuming $a\sim b \sim c$, we expect
\beq
m_u:m_c:m_t \sim \epsilon ^2: \epsilon  :1,
\label{upratios}
\eeq
where we assume,
\beq
\epsilon =  \frac{\langle \theta_{\cal U}\rangle }{\Lambda} \sim 10^{-3}.
\eeq

\subsection{The Majorana sector}
In the Majorana sector the $Z_5$ allowed leading operators are diagonal and given by,
\beq
{y_{1}\xi_1}
\frac{\theta_{\cal U}^4}{\Lambda^4}\frac{\overline{{\cal H}}_{\cal U}\overline{{\cal H}}_{\cal U}}{\Lambda_R^2}
{\cal U}^c_1{\cal U}^c_1
+{y_{2}\xi_2}
\frac{\theta_{\cal U}^2}{\Lambda^2}\frac{\overline{{\cal H}}_{\cal U}\overline{{\cal H}}_{\cal U}}{\Lambda_R^2}
{\cal U}^c_2{\cal U}^c_2
+{y_{3}\xi_3}\frac{\overline{{\cal H}}_{\cal U}\overline{{\cal H}}_{\cal U}}{\Lambda_R^2}
{\cal U}^c_3{\cal U}^c_3,
\label{Maj2}
\eeq
where $\xi_j$ are three singlets under both $A_4$ and the Pati-Salam group.
The operators relevant for the heavy Majorana mass matrix $M_R$, emerging from Eq.~\ref{Maj2},
  can be written in a more suggestive notation as,
\begin{equation}
y'_1\epsilon^4 \xi_{\rm atm} N^{c\ 2}_{\rm atm}
 + y'_2\epsilon^2 \xi_{\rm sol} N^{c\ 2}_{\rm sol}  + y'_3\xi_{\rm dec} N^{c\ 2}_{\rm dec} \;,
\label{MRops}
\end{equation}
where we have written
$\xi_{\rm atm}\equiv \xi_1$, $\xi_{\rm sol}\equiv \xi_2$, $\xi_{\rm dec}\equiv \xi_3$
and $y'_i=y_i \frac{\vev{\overline{{\cal H}}_{N^c}}^2}{\Lambda_R^2}$,
leading to a diagonal right-handed neutrino mass matrix,
\begin{equation} \label{MR0}
 M_R  =  \begin{pmatrix} \epsilon^4\tilde{M}_{1} & 0 &0 \\ 0 & \epsilon^2\tilde{M}_{2} &0 \\  0& 0& \tilde{M}_{3}\end{pmatrix} \;,
 \label{MR}
\end{equation}
where,
\beq
\tilde{M}_1=y'_1\vev{\xi_{\rm atm}} , \ \ 
\tilde{M}_2=y'_2\vev{\xi_{\rm sol}}, \ \ 
\tilde{M}_3=y'_3\vev{\xi_{\rm dec}}.
\label{RHN}
\eeq
Assuming roughly equal VEVs for $\xi_i$ we expect $\tilde{M}_1\sim \tilde{M}_2\sim \tilde{M}_3$ and hence
a very strong hierarchy of right-handed neutrino masses,
being roughly proportional to the squares of up-type quark masses in Eq.\ref{upratios}, 
hence given by the order of magnitude ratios $10^{-12}:10^{-6}:1$.
According to Eq.\ref{eq:Ynuu} the model equates up-type quark masses with Dirac neutrino masses,
apart from the Clebsch factor of 1/3 for the third family,
\footnote{Alternatively with a Clebsch factor of 3 in the second column, as discussed in the previous footnote,
we could have $m^D_{\nu2}=3m_c$.}
\beq
m^D_{\nu1}=m_u, \ \ m^D_{\nu2}=m_c, \ \ m^D_{\nu3}=\frac{m_t}{3}.
\eeq
The discrete charges (and hence powers of $\theta_{\cal U}$ and $\epsilon$) cancel in the see-saw mechanism.
This cancellation is natural, being controlled by the $Z_5$ family symmetry,
leading to the physical neutrino masses being not very hierarchical,
apart from $m_1$ which is suppressed by a factor of 9. The model therefore predicts  
a normal mass hierarchy, $m_1\ll m_2<m_3$ corresponding to 
\beq
\frac{(m^D_{\nu3})^2}{M_3} \ll 
\frac{(m^D_{\nu2})^2}{M_2} <
\frac{(m^D_{\nu1})^2}{M_1} .
\eeq
For example, $m_1\sim (m^D_{\nu3})^2/M_3 \sim  m_t^2/{(9M_3)}\sim 0.3$ meV requires $M_3 \sim 10^{16}$ GeV and hence 
$M_1 \sim 10$ TeV, $M_2 \sim 10^{10}$ GeV. 
The lightest right-handed neutrino will be difficult to observe
at colliders, due to its high mass and small Yukawa coupling of about
$10^{-6}$.
It is cosmologically unstable, decaying promptly into a neutrino plus Higgs.
Note that we identify $m_1\equiv m_{\rm dec}$, $m_2\equiv m_{\rm sol}$,
$m_3\equiv m_{\rm atm}$ and hence the heaviest right-handed neutrino of mass 
$M_3$ (from the top quark multiplet) is identified as the decoupled one $N_{\rm dec}$. 
The intermediate one of mass
$M_2$ (from the charm quark multiplet) is denoted as $N_{\rm sol}$, since it is responsible for the solar neutrino mass. The lightest right-handed neutrino of mass
$M_1$ (from the up quark multiplet) is denoted as $N_{\rm atm}$ since it is responsible for the atmospheric neutrino mass.
These identifications, familiar from SD \cite{King:1998jw}, were depicted in Fig.~\ref{421fig}.

\section{Leading Order Results}
\label{leading}
\subsection{Overview}
It is convenient to collect in one place all the lowest order quark and lepton Yukawa matrices
(in LR convention) and heavy Majorana mass matrix $M_R$
which are predicted by the model just below the high energy Pati-Salam breaking scale $\sim $ few $\times 10^{16}$ GeV,
\begin{equation} \label{Yed2}
Y^d =   \begin{pmatrix}  y_d & 0  & 0  \\ 0 & y_s & 0\\ 0  & 0 & y_b\end{pmatrix},
\ \ \ \ 
Y^e =  \begin{pmatrix}  y_d/3 & 0  & 0  \\ 0 & 3y_s & 0\\ 0  & 0 & y_b\end{pmatrix},
\end{equation}
\begin{equation} \label{Yunu2}
 Y^u =  \begin{pmatrix}  0 & b\epsilon  & 0  \\ 
a\epsilon^2 & 4b\epsilon & 0\\  a\epsilon^2 & 2b\epsilon & c\end{pmatrix},\ \ 
Y^{\nu} = \begin{pmatrix}  0 & b\epsilon  & 0  \\ 
a\epsilon^2 & 4b\epsilon & 0\\  a\epsilon^2 & 2b\epsilon & c/3\end{pmatrix},
\ \ 
 M_R  =  \begin{pmatrix} \epsilon^4\tilde{M}_{1} & 0 &0 \\ 0 & \epsilon^2\tilde{M}_{2} &0 \\  0& 0& \tilde{M}_{3}\end{pmatrix} 
\end{equation}
where we assume the phenomenologically required values of $y_d,y_s,y_b$ and we fix
$\epsilon = 10^{-3}$, which implies that the remaining parameters take natural values,
\beq
a\sim b \sim c \sim 1,   \ \ \ \ 
 \tilde{M}_1\sim \tilde{M}_2\sim \tilde{M}_3\sim 10^{16} \ GeV,
\eeq
where we allow these parameters to differ from each other by up to an order of magnitude.
The main results follow directly from the simple forms of matrices above:
\begin{itemize}
\item 
$m_e=\frac{m_d}{3}, \ \ m_{\mu}=3m_s, \ \ m_{\tau}=m_b$ ($y_d, y_s, y_b$ chosen to fit the down quark masses)
\item
$m^D_{\nu1}=m_u= |a| v_u  \epsilon^2/ \sqrt{17}, \ \ m^D_{\nu2}=m_c=\sqrt{17} |b| v_u \epsilon, \ \ 
m^D_{\nu3}=m_t/3 = |c| v_u/3$ 
\item
$M_1:M_2:M_3\sim m_u^2:m_c^2:m_t^2$ (RH neutrino masses are very hierarchical)
\item
For example, $M_1 \sim 10$ TeV, $M_2 \sim 10^{10}$ GeV, 
$M_3 \sim 10^{16}$ GeV
\item
The model predicts a normal neutrino hierarchy, due to the 
Clebsch suppression factor of 1/3 in the neutrino Yukawa mass which implies
$\frac{(m^D_{\nu3})^2}{M_3} \ll 
\frac{(m^D_{\nu2})^2}{M_2} ,
\frac{(m^D_{\nu1})^2}{M_1} $
\item
For example,
$m_1 \sim 0.3$ meV,
$m_2 \sim 8.5$ meV,
$m_3 \sim 50$ meV
(normal hierarchy)
\item
$Y^{\nu} \sim Y^u$ is the only non-diagonal matrix is responsible for all quark and lepton mixing,
which is fully specified once $a,b,c$ are fixed by up quark masses
\item
Lepton mixing angles and CP violation are predicted for the phenomenological
range of $m_2/m_3$, assuming 
a relative phase of $2\pi/5$ between the first and second columns.
\item 
The Cabibbo angle is predicted to be $\theta_C \approx 1/4$ 
or $\theta_C \approx 14^{\circ}$ at leading order
\item
The other quark mixing angles and CP violating phase are zero at leading order
\end{itemize}
The first set of relations (which are valid at the Pati-Salam breaking scale) 
are just the usual Georgi-Jarlskog (GJ) relations from $SU(5)$ \cite{Georgi:1979df}.
The tetra-model also yields an
$SO(10)$-like pattern of Dirac and heavy Majorana neutrino masses
widely studied in the literature \cite{Branco:2002kt}.
However
the light physical Majorana neutrino masses are not so hierarchical since 
the powers of $\epsilon$ cancel in the see-saw mechanism.
It has recently been shown that the serious difficulties facing thermal leptogenesis in 
$SO(10)$-like models may be circumvented when the production from the next-to-lightest right-handed neutrinos and flavour effects are properly taken into account \cite{DiBari:2010ux}, so the prospects for thermal leptogenesis
in the tetra-model look promising. Note that if we were to have $m^D_{\nu2}=3m_c$, as is possible in the alternative model discussed in the previous footnotes,
then this would increase $M_2$ by a factor of 9, enhancing the leptogenesis asymmetry 
from the next-to-lightest right-handed neutrino.
Finally, it is noteworthy that the Cabibbo angle is successfully predicted at leading order (to within one degree) as
a consequence of the vacuum alignment and quark-lepton unification,
providing the Cabibbo connection between quark and lepton mixing.
This is one of the main successes of the model, being a consequence of the $(1,4,2)$ vacuum alignment which also successfully reproduces lepton mixing, as we now discuss.

\subsection{Leading order lepton mixing}
In this subsection we discuss the leading order predictions for PMNS mixing which arise from the vacuum alignment.

The physical effective neutrino Majorana mass matrix $m^{\nu}$ is determined
from the columns of $Y^{\nu}$ via the see-saw mechanism,
\begin{eqnarray}
m^{\nu} = - v_u^2\, Y^{\nu} M^{-1}_\mathrm{R} Y^{\nu T} \; ,
\label{seesaw}
\end{eqnarray} 
where the Majorana neutrino mass
matrix $m^\nu$, defined by 
\footnote{Note that this convention for the 
light effective Majorana neutrino mass matrix $m^{\nu}$
differs by an overall complex conjugation compared to that used in the
Mixing Parameter Tools package \cite{Antusch:2005gp}.} 
$\mathcal{L}_\nu=-\tfrac{1}{2} m^\nu \overline \nu_{\mathrm{L}} 
\nu^{c}_{\mathrm{L}}$ + h.c., is diagonalised by
\begin{eqnarray}\label{eq:DiagMnu}
U_{\nu_\mathrm{L}} \,m^\nu\,U^T_{\nu_\mathrm{L}} =
\left(\begin{array}{ccc}
\!m_1&0&0\!\\
\!0&m_2&0\!\\
\!0&0&m_3\!
\end{array}
\right)\! .
\end{eqnarray}  
The PMNS matrix is then given by
\begin{eqnarray}
U_{\mathrm{PMNS}} = U_{e_\mathrm{L}} U^\dagger_{\nu_\mathrm{L}}\; .
\end{eqnarray}
We use a standard parameterization 
$
U_{\mathrm{PMNS}} = R^l_{23} U^l_{13} R^l_{12} P^l
$ 
in terms of $s^l_{ij}=\sin (\theta^l_{ij})$,
$c^l_{ij}=\cos(\theta^l_{ij})$, the Dirac CP violating phase $\delta^l$ and
further Majorana phases contained in $P^l={\rm diag}(e^{i\frac{\beta^l_1}{2}},e^{i\frac{\beta^l_2}{2}},1)$.
The standard PDG parameterization  \cite{PDG} differs slightly due to the definition of Majorana phases which are by given by $P^l_{\rm PDG}={\rm diag}(1,e^{i\frac{\alpha_{21}}{2}},e^{i\frac{\alpha_{31}}{2}})$.
Evidently the PDG Majorana 
phases are related to those in our convention by $\alpha_{21}=\beta_2^l-\beta_1^l$ and $\alpha_{31}=-\beta_1^l$,
after an overall unphysical phase is absorbed by $U_{e_\mathrm{L}}$.

Using the see-saw formula in Eq.\ref{seesaw}, with the neutrino Yukawa matrix $Y^{\nu}$ in
Eq.\ref{eq:Ynuu} and the right-handed Majorana mass matrix $M_R$ in Eq.\ref{MR0},
we find the neutrino mass matrix $m^{\nu}$, 
up to an overall irrelevant phase which may be taken to be real, can be written as 
\beq
m^{\nu} 
= m_a \begin{pmatrix} 0 & 0 & 0 \\ 0 & 1 & 1 \\ 0 & 1 & 1 \end{pmatrix} 
+ m_be^{2i\eta}\begin{pmatrix} 1 & 4 & 2 \\ 4 & 16 & 8 \\ 2 & 8 & 4  \end{pmatrix}
+m_c \begin{pmatrix} 0 & 0 & 0 \\ 0 & 0 & 0 \\ 0 & 0 & 1 \end{pmatrix} 
\label{seesaw3}
\eeq
where $m_a=|a|^2v_u^2/  |\tilde{M}_1|$, $m_b=|b|^2v_u^2/ |\tilde{M}_2|$, $m_c=|c|^2v_u^2/ (9|\tilde{M}_3|)$
are real parameter combinations which determine the three physical neutrino masses 
$m_3,m_2,m_1$, respectively. Note that $m_1$ is suppressed by a factor of 9 compared to the
other neutrino masses due to the Clebsch-Gordan factor of 1/3 in the third family Dirac neutrino mass.
We written the relative phase difference between the first two two terms as $2\eta $.
As shown recently \cite{King:2013xba}, fixing $\eta =\pm 2\pi /5$, using the phases of the singlet flavon
VEVs $\langle \xi_i \rangle$,
then determines all the lepton mixing angles and phases in terms of 
the ratio $\epsilon_{\nu}=m_b/m_a$. 
Changing the sign of the phase $\eta =\pm 2\pi /5$ leaves
the predictions for the angles unchanged, but reverses the signs of the Dirac and Majorana
phases \cite{King:2013iva}. 
Here we shall select $\eta =2\pi /5$ since it leads to a negative Dirac phase, preferred by the
most recent global fits \cite{GonzalezGarcia:2012sz}.
Since $\eta$ is crucial to the predictions in the lepton sector,
it is worthwhile discussing the origin of this phase in more detail.

In order to understand the origin of phases which enter the neutrino mass matrix $m^{\nu}$, it is worth recalling that 
the operators responsible for the neutrino Yukawa and Majorana masses are those given in
Eqs.\ref{ynuops} and \ref{MRops}.
 Implementing the see-saw mechanism, the effective neutrino mass matrix 
 $m^{\nu}$ in Eq.~\ref{seesaw3} emerges from the flavon combinations,
\beq
\label{seesaw2}
m^{\nu}\sim  \frac{\vev{\phi_{\rm atm}}\vev{\phi_{\rm atm}}^T}{\langle \xi_{\rm atm} \rangle}
+  \frac{\vev{\phi_{\rm sol}}\vev{\phi_{\rm sol}}^T}{\langle \xi_{\rm sol} \rangle}
+  \frac{\vev{\phi_{\rm dec}}\vev{\phi_{\rm dec}}^T}{\langle \xi_{\rm dec} \rangle}.
\eeq
Notice that the powers of $\epsilon$ cancel in the see-saw mechanism, leading to a rather 
mild hierarchy in the neutrino sector.
Since we are assuming that the original theory respects CP, the only source of phases can be the VEVs
of flavons. The phase $\eta =2\pi /5$ then must arise from the difference between flavon VEVs.
The phases of flavon VEVs arise in the context of spontaneous CP violation from discrete symmetries as discussed in 
\cite{Antusch:2013wn}, and we shall follow the strategy outlined there.
The basic idea is to impose CP conservation on the theory so that all couplings and masses are real.
Note that the $A_4$ assignments in Table~\ref{A4PS} do not involve the complex
singlets $1',1''$ or any complex Clebsch-Gordan coefficients so that the definition of CP is 
straightforward in this model and hence all the different ways 
that CP may be defined in $A_4$ \cite{Ding:2013bpa} are equivalent for our purposes 
(see \cite{Antusch:2013wn} for a discussion of this point).
The CP symmetry is broken in a discrete way by the form of the superpotential terms.

We have already stated that the flavon VEVs 
$\vev{\phi_{\rm atm}}$ and $\vev{\phi_{\rm sol}}$ are real and in this case the phase
$\eta$ must arise from the singlet flavons VEVs $\langle \xi_i \rangle$.
For example, Eq.~\ref{seesaw2} shows that the phase $\eta$ in Eq.\ref{seesaw3}
could originate from the solar flavon VEV
$\langle \xi_{\rm sol} \rangle \sim e^{-4i\pi/5}$, if the atmospheric
flavon vev $\langle \xi_{\rm atm} \rangle $ is real and positive. 
This can be arranged if the right-handed neutrino flavon vevs arise from 
$Z_5$ invariant quintic terms in the superpotential,
\be
 g_1 P_1\left(\frac{\xi_{\rm atm}^5}{\Lambda_1^3}  -\mu_1^2\right) + 
g_2P_2 \left(\frac{\xi_{\rm sol}^5}{\Lambda_2^3}  - \mu_2^2\right)
+ 
g_3P_3 \left(\frac{\xi_{\rm dec}^5}{\Lambda_3^3}  - \mu_3^2\right) ,
\label{Rflavon}
\ee
where, as in \cite{Antusch:2013wn}, the driving singlet fields $P_i$ denote linear combinations 
of identical singlets and all couplings and masses are real due to CP conservation.
The F-term conditions from Eq.\ref{Rflavon} are,
\begin{equation}
 \left| \frac{\langle \xi_{\rm atm} \rangle^5}{\Lambda_1^3} - \mu_1^2\right|^2 
 = \left| \frac{\langle \xi_{\rm sol} \rangle^5}{\Lambda_2^3} - \mu_2^2\right|^2 
  = \left| \frac{\langle \xi_{\rm dec} \rangle^5}{\Lambda_3^3} - \mu_3^2\right|^2=  0 .
\end{equation} 
These are satisfied, for example, by $\langle \xi_{\rm atm} \rangle = |(\Lambda_1^3 \mu_1^2)^{1/5}|$ and 
$\langle \xi_{\rm sol} \rangle =  |(\Lambda_2^3\mu_2^2)^{1/5}|e^{4i\pi/5}$
and $\langle \xi_{\rm dec} \rangle = |(\Lambda_3^3 \mu_3^2)^{1/5}|$
where we arbitrarily select two of the phases to be 
zero and the solar phase to be $-4\pi /5$ from amongst a discrete set of possible choices
in each case. More generally we select a phase difference of $-4\pi /5$ between
$\langle \xi_{\rm atm} \rangle$ and $\langle \xi_{\rm sol} \rangle$, 
with an arbitrary phase for $\langle \xi_{\rm dec} \rangle$,
since the overall phase is not physically
relevant and the decoupled phase is not important, which would happen one in five times by chance.
In the basis where the right-handed neutrino masses are real and positive
this is equivalent to having a phase $\eta  = 2\pi /5$ in Eq.~\ref{seesaw3}.

\begin{table}
	\centering
		\begin{tabular}{|c||c|c|c|c|c|c|c|c|c|}
			\hline
			 $\epsilon_{\nu}$ & $m_2/m_3$ &  $\theta^l_{12}$ 
			 & $\theta^l_{13}$  & $\theta^l_{23}$  & $\delta^l$  & $\beta^l_1$ & $\beta^l_2$ & $\alpha_{21}$ & $\alpha_{31}$ \\ \hline \hline
			0.057 &0.166 &34.3$^{\circ}$ & 8.75$^{\circ}$ & 39.6$^{\circ}$ & 258$^{\circ}$ &323$^{\circ}$& 77.5$^{\circ}$&114$^{\circ}$&37$^{\circ}$\\ \hline   
			0.058 &0.170 &34.2$^{\circ}$ & 8.9$^{\circ}$ & 39.7$^{\circ}$ & 259$^{\circ}$ &322$^{\circ}$& 76$^{\circ}$&114$^{\circ}$&37.5$^{\circ}$\\ \hline 
			0.059 &0.174 &34.1$^{\circ}$ & 9.1$^{\circ}$ & 39.8$^{\circ}$ & 260$^{\circ}$ &322$^{\circ}$& 75$^{\circ}$&113$^{\circ}$&38$^{\circ}$\\ \hline
			0.060 &0.178 &34.0$^{\circ}$ & 9.3$^{\circ}$ & 39.9$^{\circ}$ & 260.5$^{\circ}$ &321$^{\circ}$& 73$^{\circ}$&112$^{\circ}$&39$^{\circ}$\\ \hline  
			0.061 &0.182 &33.9$^{\circ}$ & 9.4$^{\circ}$ & 40.0$^{\circ}$ & 261$^{\circ}$ &320$^{\circ}$& 72$^{\circ}$&112$^{\circ}$&40$^{\circ}$\\ \hline     
				\end{tabular} 
			\caption{The leading order predictions for PMNS parameters as a function of 
			$\epsilon_{\nu}=m_b/m_a$ and hence $m_2/m_3$,
			for $m_1=0.3$ meV and $m_2=50$ meV.
			Note that these predictions assume $\eta = 2\pi/5$.
			The predictions are obtained numerically using the Mixing Parameter Tools (MPT)
package based on \cite{Antusch:2005gp}, taking into account the different conventions.
The last two columns also show the PDG Majorana phases  \cite{PDG} given by 
$\alpha_{21}=\beta_2^l-\beta_1^l$ and $\alpha_{31}=-\beta_1^l$.
			}
		\label{predictions142}
\end{table}

Returning to Eq.~\ref{seesaw3}, with $\eta =2\pi /5$,
the six predictions vary with $\epsilon_{\nu}$, or equivalently $m_2/m_3$,
 as shown in Table~\ref{predictions142} for a fixed value of $m_1=0.3$ meV.
It is remarkable that, for the physical range of $m_2/m_3$, the PMNS lepton mixing angles are predicted to be
$\theta^{l}_{12}\approx 34^{\circ}$, $\theta^{l}_{23}\approx 40^{\circ}$ and  
$\theta^{l}_{13}\approx 9^{\circ}$,
which agree with the current best fit values for a normal neutrino mass hierarchy,
together with the CP violating oscillation phase $\delta^{l} \approx 260^\circ$
and Majorana phases $\beta^l_1 \approx 322^\circ$ and $\beta^l_2 \approx 75^\circ$
corresponding to the PDG Majorana phases $\alpha_{21} \approx 113^\circ$ and $\alpha_{31} \approx 38^\circ$.
\footnote{
If we were to set $m_1=0$ and choose the opposite phase $\eta =-2\pi /5$ then 
we would find the results presented previously in \cite{King:2013xba}, namely
 $\theta^{l}_{12}\approx 34^{\circ}$, $\theta^{l}_{23}\approx 41^{\circ}$,
$\theta^{l}_{13}\approx 9.5^{\circ}$, $\delta^{l} \approx 106^\circ$.
Note that the presence of the non-zero mass $m_1=0.3$ meV reduces the reactor angle by about half a degree,
bringing it even closer to the central value observed by Daya Bay of $\theta^{l}_{13}\approx 8.7^{\circ}$
\cite{DayaBay}.
Such $m_1$ corrections to SD were first considered in \cite{Antusch:2010tf}.}
We emphasise that the tetra-model predicts both a normal hierarchy and an atmospheric angle in the first octant.
Both these predictions will be subjected to experimental scrutiny in the near future
\cite{Winter:2013ema}.

The neutrinoless double beta decay ($0\nu \beta \beta$) parameter $|m_{ee}|$, may be estimated
using the standard PDG formula \cite{PDG}. For the parameters in Table~\ref{predictions142} we find 
$|m_{ee}|\approx 1.5$ meV, below the sensitivity of most planned $0\nu \beta \beta$ experiments,
as expected for such a hierarchical neutrino mass pattern.
If the lightest neutrino mass were artificially increased by an order of magnitude to 
$m_1=3$ meV, and the other parameters unchanged, we would find $|m_{ee}|\approx 2.4$ meV.
This demonstrates the insensitivity of $|m_{ee}|$ to the lightest neutrino mass and shows that, 
although significant cancellations could in principle occur in the calculation of $|m_{ee}|$
for a normal hierarchy
\cite{King:2013psa}, for the predicted PMNS parameters of the tetra-model such cancellations do not occur.

\section{Higher Order Corrections }
\label{higher}
\subsection{Higher Order Operators}
Since the vacuum alignments are achieved by a renormalisable superpotential, it is possible
that the HO corrections to vacuum alignment, originating from non-renormalisable operators,
are highly suppressed compared to the LO alignments. We are free to assume this, since the messenger scale
associated with such HO operators is unconstrained by the model. Therefore we shall ignore the corrections to vacuum alignment in our analysis.

The HO operators in the down Yukawa sector arise from cubic insertions flavon fields
$ \phi_{{\cal D}^c_j}^3 $ which are singlets under the $Z_3$ symmetries.
These insertions are accompanied by messenger mass suppressions $\vev{\Sigma}^3$ or $\vev{\Sigma_{15}}^3$
which are also $Z_3$ singlets.
The flavons $\phi_{{\cal D}^c_i}$ may lead to significant 
suppression since the factor $\vev{\phi_{{\cal D}^c_3}}/  \vev{\Sigma}$ is responsible for the bottom quark Yukawa coupling,
and the other flavons are responsible for the
strange and down quark masses and so their contribution will be highly suppressed. 
Therefore the dominant HO correction arises from insertions
of $\phi_{{\cal D}^c_3}^3$ corresponding to a suppression of order $y_b^3$.
The most important HO operators in the down sector arising are then, dropping the coupling constants and scales,
\beq
h_d( \phi_{{\cal D}^c_3}^3\phi_{{\cal D}^c_1}{\cal Q}){\cal D}^c_1
+h_{\cal D}( \phi_{{\cal D}^c_3}^3\phi_{{\cal D}^c_2}{\cal Q}){\cal D}^c_2
+h_d( \phi_{{\cal D}^c_3}^3\phi_{{\cal D}^c_3}{\cal Q}){\cal D}^c_3,
\label{YukdHO}
\eeq
The $A_4$ contractions in the above HO operators differ from the LO contractions
previously. In particular an $A_4$ singlet is achieved by contracting $A_4$ triplets inside the round
brackets in Eq.~\ref{YukdHO} in all possible ways. 
We find that the allowed contractions from the above operators result in dominant corrections to the 
down Yukawa matrix of the form,
\begin{equation} \label{Yed}
Y^d =   \begin{pmatrix}  y_d & {\cal O}(y_sy_b^3)  & 0  \\ {\cal O}(y_dy_b^3)  & y_s & 0\\ 0  & 0 & y_b\end{pmatrix},
\end{equation}
The corrections are negligible providing $\tan \beta$ is not too large. For example, for $\tan \beta \lsim 10$,
we have $y_b \lsim 0.1$ and hence $y_b^3 \lsim 10^{-3}$, resulting in a very small and 
negligible contribution to the Cabibbo angle.

The HO operators in the up Yukawa sector arise from quintic insertions of flavon fields
$ \phi_{{\cal U}^c_j}^5 $.
These insertions are accompanied by messenger mass suppressions $\vev{\Sigma'}^5$ or $\vev{\Sigma'_{15}}^5$
which are also $Z_5$ singlets.
The flavons $\phi_{{\cal U}^c_i}$ should not lead to too much
suppression since the factor $\vev{\phi_{{\cal U}^c_3}}/  \vev{\Sigma'_{15}}$ is responsible for the top quark Yukawa coupling,
and the other flavons are assumed to have similar VEVs (with the hierarchy
in the up sector generated by powers of $\epsilon$ associated with powers of the $\theta_{\cal U}$ VEV).
For example if we were to assume that each insertion of flavon field were associated with a mild suppression factor
of 1/2 then the quintic insertions would correspond to a suppression of $1/2^5\sim 1/30$.
Such corrections may dominate over those coming from the down sector, at least for low or moderate
$\tan \beta$,
and here we assume that they 
provide the most important corrections to quark mixing.

The operator insertions $\phi_{{\cal U}^c_j }\xi_j^2$ are also singlets and should be included.
We shall assume that they are competitive with the previous operators, although the messenger masses
associated with these operators is not determined.
The most important HO operators in the up sector arising from the insertions of the above operators are then, dropping the coupling constants and scales,
\bea
&\sum_{j=1}^3 &
h_u\theta_{\cal U}^2(\phi_{{\cal U}^c_j}^5\phi_{{\cal U}^c_1}{\cal Q}){\cal U}^c_1
+h_u\theta_{\cal U}(\phi_{{\cal U}^c_j}^5\phi_{{\cal U}^c_2}{\cal Q}){\cal U}^c_2
+h_u(\phi_{{\cal U}^c_j}^5\phi_{{\cal U}^c_3}{\cal Q}){\cal U}^c_3 \nonumber \\
+ &\sum_{j=1}^3 & h_u\theta_{\cal U}^2(\phi_{{\cal U}^c_j }\xi_j^2\phi_{{\cal U}^c_1}{\cal Q}){\cal U}^c_1
+h_u\theta_{\cal U}(\phi_{{\cal U}^c_j }\xi_j^2\phi_{{\cal U}^c_2}{\cal Q}){\cal U}^c_2
+h_u(\phi_{{\cal U}^c_j }\xi_j^2\phi_{{\cal U}^c_3}{\cal Q}){\cal U}^c_3,
\label{YukuHO}
\eea
where we assume that each of these operators will lead to a correction compared to the corresponding
LO operator with an extra suppression factor of order $\sim 1/30$ according to the above
example. Note that these HO operators respect the hierarchy generated by powers
of $\epsilon$ associated with powers of the $\theta_{\cal U}$ VEV, so do not disturb the up type quark
mass hierarchy. The $A_4$ contractions in the above HO operators differ from the LO contractions
previously. In particular an $A_4$ singlet is achieved by contracting $A_4$ triplets inside the round
brackets in Eq.~\ref{YukuHO} in all possible ways. 
In this case, due to the pattern of alignments in Eq.\ref{phiu}, and the fact that all operator
insertions contribute equally, we expect a large number of allowed contractions, with similar multiplicative corrections
filling all entries of the up and neutrino Yukawa matrices. However the corrections involving ${\cal Q}_1$
(i.e. the correction in the first row of the up Yukawa matrix) may be smaller by a factor of 1/4 due to the
alignments in Eq.\ref{phiu}.  
We shall discuss the phenomenological impact of these
corrections later.

In the Majorana sector the above charge assignments allow higher order mixed terms 
such as 
\beq
\frac{1}{\Lambda^3\vev{\Sigma}\Lambda_R^2}
\theta_{\cal U}^3
(\phi_{{\cal U}^c_1}.\phi_{{\cal U}^c_2})\overline{{\cal H}}_{\cal U}\overline{{\cal H}}_{\cal U}{\cal U}^c_1{\cal U}^c_2,
\label{Maj3}
\eeq
leading to an off-diagonal right-handed neutrino mass matrix,
\begin{equation} \label{MR}
 M_R  =  \begin{pmatrix} \epsilon^4\tilde{M}_{1} &  \epsilon^3\tilde{M}_{12}  &0 \\ 
 \epsilon^3\tilde{M}_{12} & \epsilon^2\tilde{M}_{2} &0 \\  0& 0& \tilde{M}_{3}\end{pmatrix} \;,
\end{equation}
where,
\beq
\tilde{M}_{12}=y_{12}\frac{v_{{\cal U}^c_1}v_{{\cal U}^c_2}}{\vev{\Sigma}} \frac{\vev{\overline{{\cal H}}_{N^c}}^2}{\Lambda_R^2}.
\label{RHN}
\eeq
This operator contributes off-diagonal terms to the right-handed neutrino mass matrix
of a magnitude which depends on the absolute scale of the 
flavon vevs $\vev{\phi_{{\cal U}^c_1}}$ and $\vev{\phi_{{\cal U}^c_2}}$ compared to 
$\langle \xi_1 \rangle $ and $\langle \xi_2\rangle $.
If all flavon vevs and messenger scales in the neutrino sector
are set equal then we would expect $\tilde{M}_{12}\sim \tilde{M}_{1}$, with a significant contribution
to atmospheric mixing even if $\epsilon \sim 10^{-3}$ due to the hierarchical nature of the neutrino Yukawa
matrix. However this correction may be completely insignificant if 
$\vev{\phi_{{\cal U}^c_i}}\ll \vev{\xi_i}$ which would imply $\tilde{M}_{12}\ll \tilde{M}_{1}$.
Since we require $\vev{\xi_i} \sim 10^{16}$ GeV, in order to obtain a small enough value of $m_1$,
this is tantamount to assuming that $\vev{\phi_{{\cal U}^c_i}}\ll 10^{16}$ GeV.

\subsection{The up quark and neutrino Yukawa matrix at higher order}
As discussed in the previous subsection, 
the down and charged lepton Yukawa matrices receive negligible HO corrections 
and may be neglected to good approximation.
We therefore assume that the down quark Yukawa matrix 
is unchanged from its diagonal form given earlier in Eq.~\ref{Yed}.
On the other hand, the up quark and neutrino Yukawa matrices are expected to be corrected
by a complicated set of operators and contractions as shown in Eq.~\ref{YukuHO}, 
with the corrections being of order $\sim {\cal O}(1/30)$. 
This implies that CKM mixing originates entirely from the up quark Yukawa matrix which takes the general form,
in the presence of (complex) HO corrections,
\begin{equation} \label{YHO}
\begin{pmatrix}  \varepsilon_{11}\epsilon^2 & b\epsilon (1+\varepsilon_{12})  & \varepsilon_{13}c  \\ 
a\epsilon^2(1+\varepsilon_{21}) & 4b\epsilon (1+\varepsilon_{22}) &\varepsilon_{23}c\\  
a\epsilon^2(1+\varepsilon_{31}) & 2b\epsilon (1+\varepsilon_{32}) & c(1+\varepsilon_{33})\end{pmatrix},
\end{equation}
with a similar matrix for the neutrino Yukawa matrix, differing only by Clebsch factors.

Each of the parameters $\varepsilon_{ij}$ may in general receive contributions from 
several operator contractions arising from Eq.~\ref{YukuHO}, each with a quantised phase (a fifth root of unity)
and each entering with an arbitrary coefficient. 
The parameters $\varepsilon_{ij}$ are therefore taken to be arbitrary complex 
parameters, with modulus $\lsim {\cal O}(1/30)$, which correct the leading order 
mixing predictions in both the lepton and quark sectors. 
In the limit $\varepsilon_{ij}=0$ the matrices reduce to 
the simple forms in Eq.\ref{Yunu2}.
Before discussing the effect of the higher order corrections in detail, it is useful to begin with an overview of 
the significance of the three columns of 
this matrix for lepton and quark mixing as follows:
\begin{itemize}
\item 
The first column of Eq.\ref{YHO} is mainly responsible for the atmospheric neutrino mass and mixing.
The reactor angle and leptonic CP violation originates
from the interplay between the first and second columns,
being sensitive to the relative phase between these two columns.
The first column is irrelevant to CKM to good approximation, being suppressed
by $\epsilon^2$ which is related to the smallness of the up quark mass.

\item
The second column of Eq.\ref{YHO} is mainly responsible for the solar neutrino mass and mixing.
The second column is also responsible for the Cabibbo angle,
providing the Cabibbo connection
between quark and lepton mixing.
We saw that the Cabibbo angle is given at LO by 
$\theta_C \approx 1/4$, however the HO corrections will modify this prediction,
along with the PMNS predictions.

\item
The third column of Eq.\ref{YHO} is approximately decoupled from the see-saw mechanism 
due to the smallness of $m_1$ (the SD mechanism) and so is unimportant for lepton mixing.
However the third column is responsible for the small quark mixing angles and quark CP violation.
It is also responsible for the top quark Yukawa coupling.

\end{itemize}
\subsection{Higher order corrections to quark mixing}
The up quark Yukawa matrix defined by 
\footnote{Note that this convention for the quark Yukawa matrix
differs by an Hermitian conjugation compared to that used in the
Mixing Parameter Tools package \cite{Antusch:2005gp} due to the RL convention used there.} 
$\mathcal{L}=-v^uY^u_{ij}\overline u^i_{\mathrm{L}} u^j_{\mathrm{R}}$ + h.c.  
is diagonalised by,
\begin{eqnarray}\label{DiagYu}
U_{u_\mathrm{L}} \, Y^u \,U^\dagger_{u_\mathrm{R}} =
\left(\begin{array}{ccc}
\!y_{u}&0&0\!\\
\!0&y_{c}&0\!\\
\!0&0&y_{t}\!
\end{array}
\right)\! .
\end{eqnarray}  
The CKM matrix is given by
\begin{eqnarray}
U_{\mathrm{CKM}} = U_{u_\mathrm{L}} U^\dagger_{d_\mathrm{L}}\; ,
\end{eqnarray}
where $U_{d_\mathrm{L}}$ is a diagonal matrix of phases since $Y^d$ is diagonal.
We use the PDG parameterization in the standard notation 
$U_{\mathrm{CKM}} = R^q_{23} U^q_{13} R^q_{12}$
in terms of $s^q_{ij}=\sin (\theta^q_{ij})$ and 
$c^q_{ij}=\cos(\theta^q_{ij})$ and the CP violating phase $\delta^q$. 
Since the down Yukawa matrix is diagonal, 
the CKM matrix is given by $U_\mathrm{CKM}=U_{u_{\mathrm{L}}}\!\!\cdot
\mbox{diag}\,(1,e^{i \beta_2},e^{i \beta_3})$.
The hierarchical form of the columns of $Y^u$,
\begin{equation} \label{YuHO}
Y^u =  \begin{pmatrix}  \varepsilon_{11}\epsilon^2 & b\epsilon (1+\varepsilon_{12})  & \varepsilon_{13}c  \\ 
a\epsilon^2(1+\varepsilon_{21}) & 4b\epsilon (1+\varepsilon_{22}) &\varepsilon_{23}c\\  
a\epsilon^2(1+\varepsilon_{31}) & 2b\epsilon (1+\varepsilon_{32}) & c(1+\varepsilon_{33})\end{pmatrix}
\equiv
 \left(
\begin{array}{ccc}
d & p & s \\
e & q & t \\
f & r & u
\end{array}
\right),
\end{equation}
implies that $U_{u_{\mathrm{L}}}$
is determined by, 
\begin{eqnarray}\label{eq:DiagYe}
U_{u_{\mathrm{L}}} \cdot 
 \left(
\begin{array}{ccc}
d & p & s \\
e & q & t \\
f & r & u
\end{array}
\right)
=\left(\begin{array}{ccc}
* & 0 & 0 \\
* & * & 0 \\
* & * & *
\end{array}\right).
\end{eqnarray} 
This is the same procedure that was followed for right-handed charged lepton sequential dominance
\cite{Antusch:2004re}. Indeed here we have an analogous right-handed up-quark sequential dominance,
with the third right-handed up quark dominating over the second, which in turn dominates over the first in their 
contributions to the up quark Yukawa matrix in Eq~\ref{YuHO}.
We hence obtain for the CKM parameters,
writing $t^q_{ij}=\tan (\theta^q_{ij})$,   
\begin{subequations}\label{mixings}
\begin{eqnarray}
\label{q12} 
e^{i \beta_2} t^q_{12} &\approx&  \frac{-\frac{s}{u}+\frac{p}{r}}{\frac{t}{u}-\frac{q}{r}} 
\approx - \frac{1}{4}\left(1+ \varepsilon_{12}-\varepsilon_{22}+\varepsilon_{23}/2-2\varepsilon_{13}   \right)
   \; ,\\
\label{q23}
e^{-i \delta^q}e^{i \beta_3} s^q_{13} &\approx&  
- \frac{(  c^q_{12} \, s + s^q_{12}e^{i \beta_2} \,t )}{u}
\approx - \left(c^q_{12}\varepsilon_{13}+s^q_{12}e^{i \beta_2}\varepsilon_{23}\right)
 \;,\\
\label{q13}
e^{i \beta_3} t^q_{23}  &\approx& 
 \frac{ s^q_{12} \, s - c^q_{12}e^{i \beta_2} \,t }{c^q_{13} u} 
 \approx \left(s^q_{12}\varepsilon_{13}-c^q_{12}e^{i \beta_2}\varepsilon_{23}\right)
\end{eqnarray}\end{subequations}
The parameters $\varepsilon_{ij}$ are complex and the phases on the LHS of the above equations
are fixed by the requirement that the mixing angles are real and positive.
We have checked that these results very accurately reproduce the numerical results from the 
MPT package \cite{Antusch:2005gp}, to within
an accuracy of better than 0.1\% (taking into account the different conventions used there).

From the above results we find the simpler but less accurate approximations:
\begin{subequations}\label{mixings}
\begin{eqnarray}
\label{q120} 
\theta^q_{12}&\approx&
 \frac{1}{4}|1+ \varepsilon_{12}-\varepsilon_{22}|
   \; ,\\
\label{q230}
\theta^q_{23}&\approx& |\varepsilon_{23}|
 \;,\\
\label{q130}
\theta^q_{13}&\approx& |\varepsilon_{23}/4-\varepsilon_{13}|
 \;,\\
 \label{phase}
\frac{\varepsilon_{13}}{\varepsilon_{23}}
&\approx& t^q_{12}-\frac{s^q_{13}}{t^q_{23}c^q_{12}}e^{-i \delta^q}
\end{eqnarray}\end{subequations}

Hence we find the following estimates:
\begin{itemize}
\item
From \ref {q120}, the Cabibbo angle requires $|\varepsilon_{12}-\varepsilon_{22}|\sim 0.07 \sim {\cal O}(\lambda^2)$
\item
From \ref {q230}, $V_{cb}$ is determined by $|\epsilon_{23}|\sim 0.04  \sim {\cal O}(\lambda^2)$
\item
From \ref {q130}, $V_{ub}$ is determined by $|\varepsilon_{23}/4-\varepsilon_{13}| \sim {\cal O}(\lambda^3)$
\item
From \ref{phase}, the CP phase $\delta^q\sim 70^{\circ}$ requires $Arg\left(\frac{\varepsilon_{13}}{\varepsilon_{23}}\right)\sim 22^{\circ}$
and $|\frac{\varepsilon_{13}}{\varepsilon_{23}}|\sim 0.22$
\end{itemize}
The ratio $|\frac{\varepsilon_{13}}{\varepsilon_{23}}|\sim 0.22$ is close the value 
$|\frac{\varepsilon_{13}}{\varepsilon_{23}}|\sim 1/4$ expected from the vacuum alignments.

\subsection{Higher order corrections to lepton mixing}
We expect the neutrino Yukawa matrix which to have similar corrections to those previously
considered for the up quark sector. However, as already mentioned, 
the HO corrections appearing in the third column of the Yukawa matrix, in particular
$\varepsilon_{13},\varepsilon_{23}$,
which are necessary for obtaining the small quark mixing angles and quark CP violation,
will be relatively unimportant for lepton mixing. 
On the other hand,
the HO corrections appearing in the first column of the Yukawa matrix,
are unimportant for quark mixing but will affect lepton mixing.
Only the HO corrections in the second column are important for both quark and lepton mixing,
$\varepsilon_{12},\varepsilon_{22}$ are important for correcting the Cabibbo angle.

The important message from the quark sector is that one expects that all the HO corrections relevant for quark
mixing angles to be small, and so we may infer that the neutrino Yukawa matrix involves similar corrections
$|\varepsilon_{ij}|\lsim \lambda^2$,
where $\lambda=0.225$ is the Wolfenstein parameter. 
In addition the right-handed neutrino mass matrix may gain small off-diagonal
entries at HO due to the operators discussed previously, which will lead to further additional corrections
unrelated to the quark sector. However, as discussed, if flavour is broken well below the
PS breaking scale
then such Majorana corrections are negligible.
Therefore, we need only consider the effect of small corrections $|\varepsilon_{ij}|$ to the elements of the 
neutrino Yukawa matrix, with the most important corrections arising from the first two columns,
\footnote{Note that $Y^{\nu}$ is diagonalised
by $U'_{\nu_\mathrm{L}} \, Y^{\nu} \,{U'}^\dagger_{{\nu}_\mathrm{R}} $
where $U'_{\nu_\mathrm{L}}$ is not the same as $U_{\nu_\mathrm{L}}$
in Eq.\ref{eq:DiagMnu}.
In fact $U'_{\nu_\mathrm{L}}$ is rather similar (but not identical due to Clebsch factors) to $U_{u_\mathrm{L}}$ 
which diagonalises the up quark Yukawa matrix in Eq.\ref{DiagYu}. 
Therefore $U'_{\nu_\mathrm{L}}$ is 
also of similar form to the CKM matrix, $U'_{\nu_\mathrm{L}}\sim U_\mathrm{CKM}$.}

\begin{figure}
\centering
\includegraphics[width=0.49\textwidth]{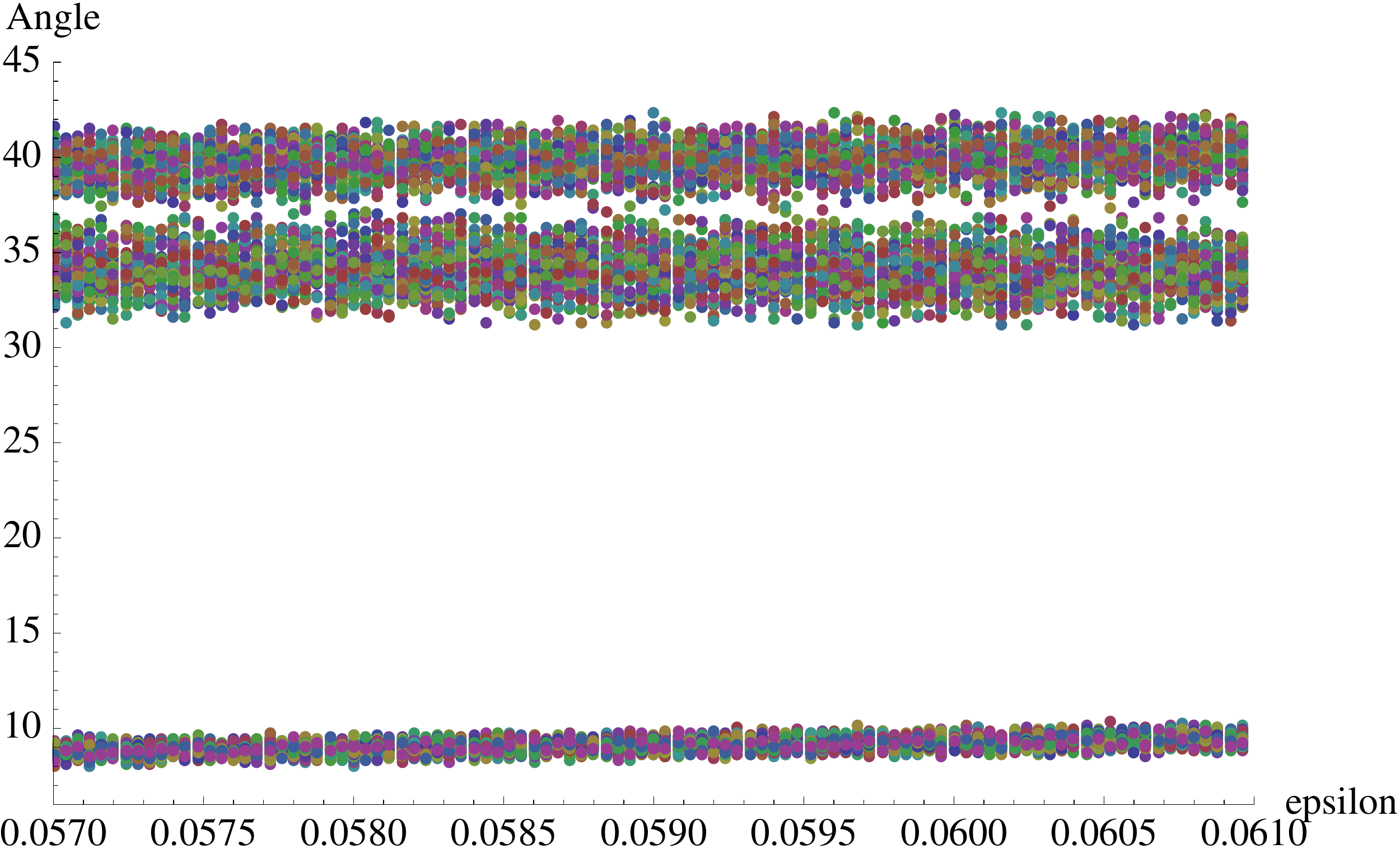}
\includegraphics[width=0.49\textwidth]{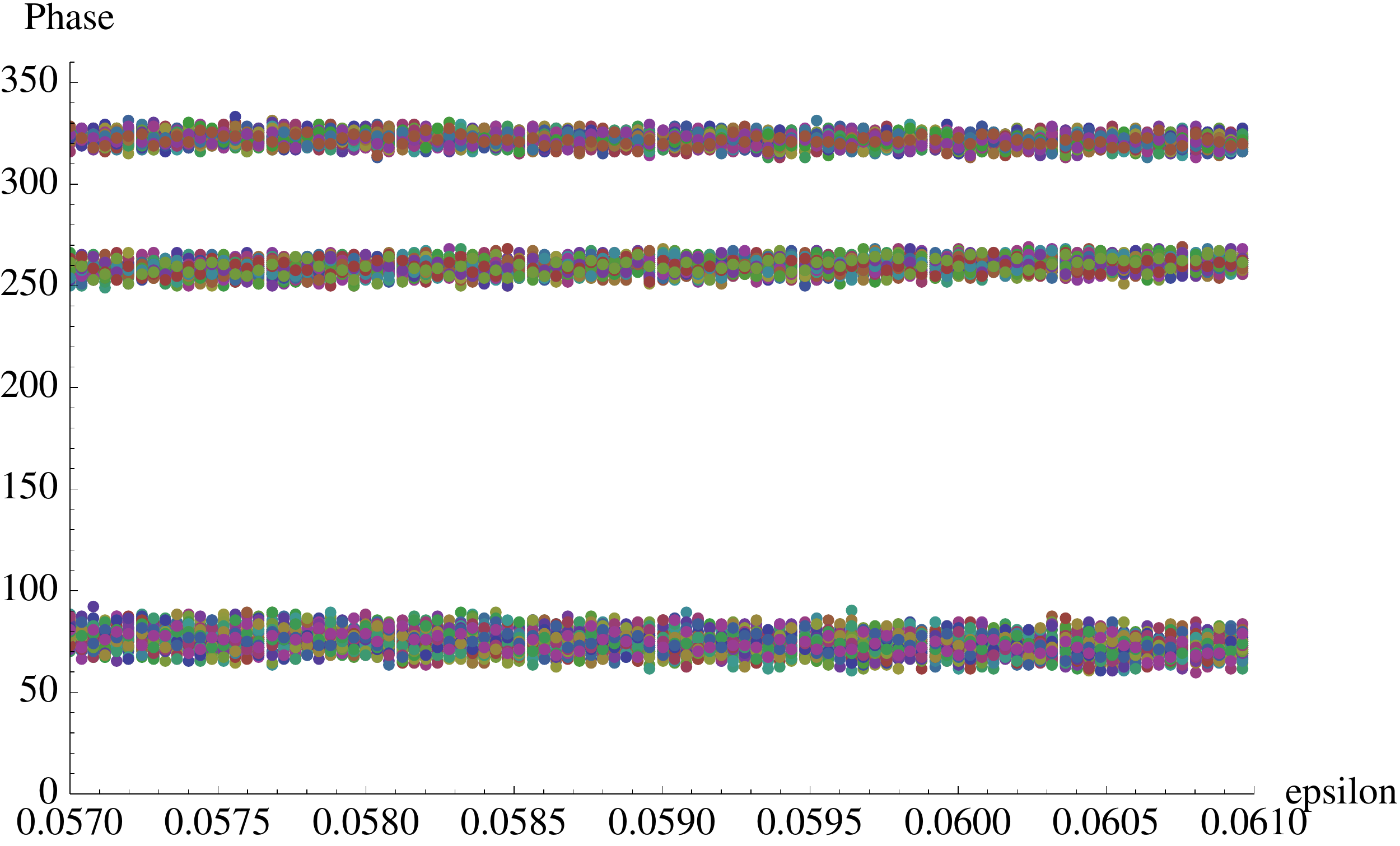}
    \caption{Overview of the PMNS predictions including the ``noise'' of the higher order corrections.
    Left panel shows the predictions for the atmospheric angle $\theta^l_{23}$ (upper) and
    solar angle $\theta^l_{12}$ (middle) and reactor angle $\theta^l_{13}$ (lower) 
    in the presence of HO corrections. Right panel shows the predictions for the oscillation phase angle $\delta^l$ 
    (middle), together with the Majorana phase $\beta_2^l$ (upper)
    and the Majorana phase $\beta_1^l$ (lower) 
    in the presence of HO corrections.
    The predictions are all given in degrees and presented as a function of 
			$\epsilon_{\nu}=m_b/m_a$ and hence $m_2/m_3$,
			for $m_1=0.3$ meV and $m_2=50$ meV.
			Note that these predictions assume $\eta = 2\pi/5$.
			The predictions are obtained numerically using the Mixing Parameter Tools (MPT)
package based on \cite{Antusch:2005gp}, taking into account the different conventions.
} \label{overview}
\vspace*{-2mm}
\end{figure}

\begin{figure}
\centering
\includegraphics[width=0.49\textwidth]{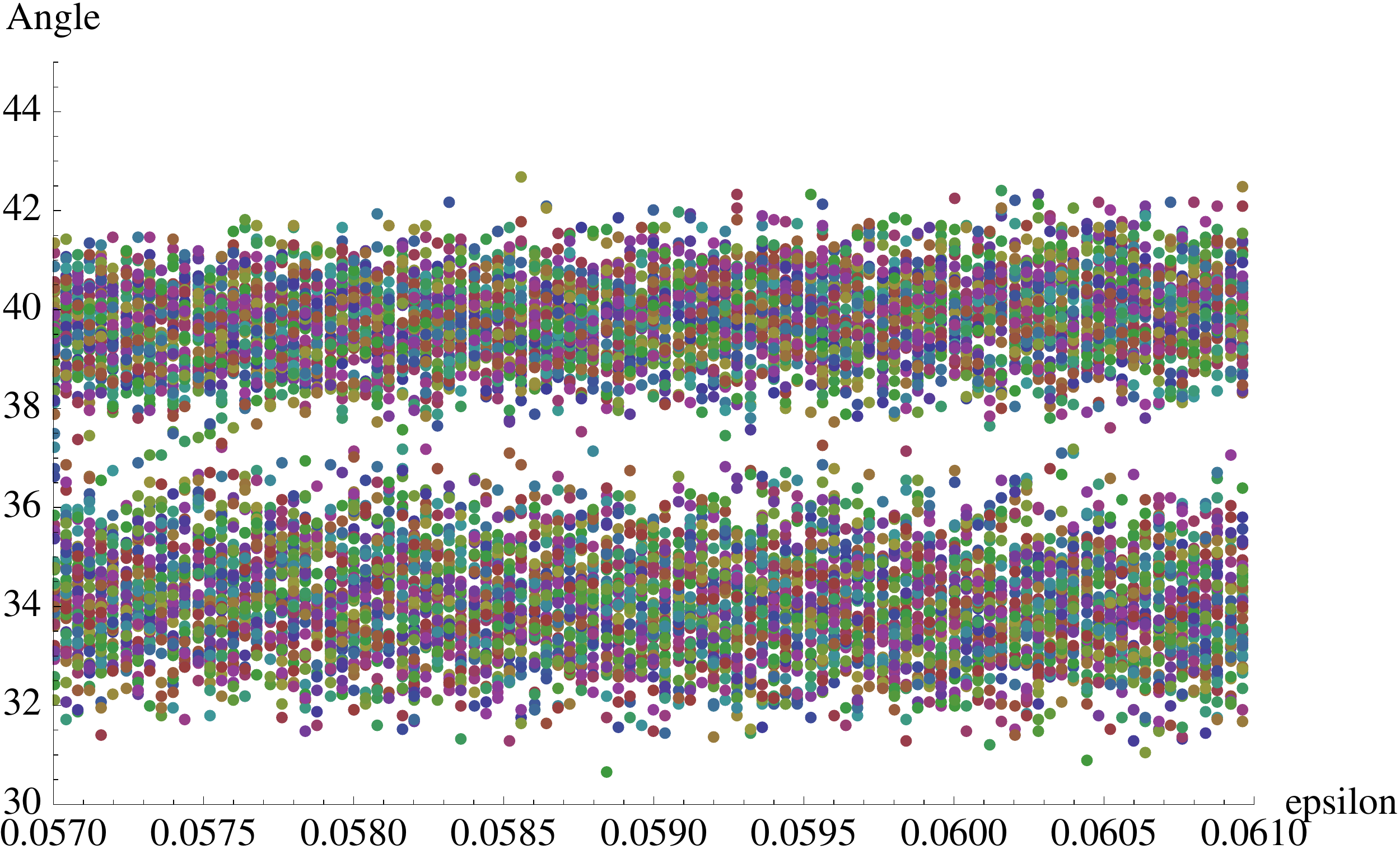}
\includegraphics[width=0.49\textwidth]{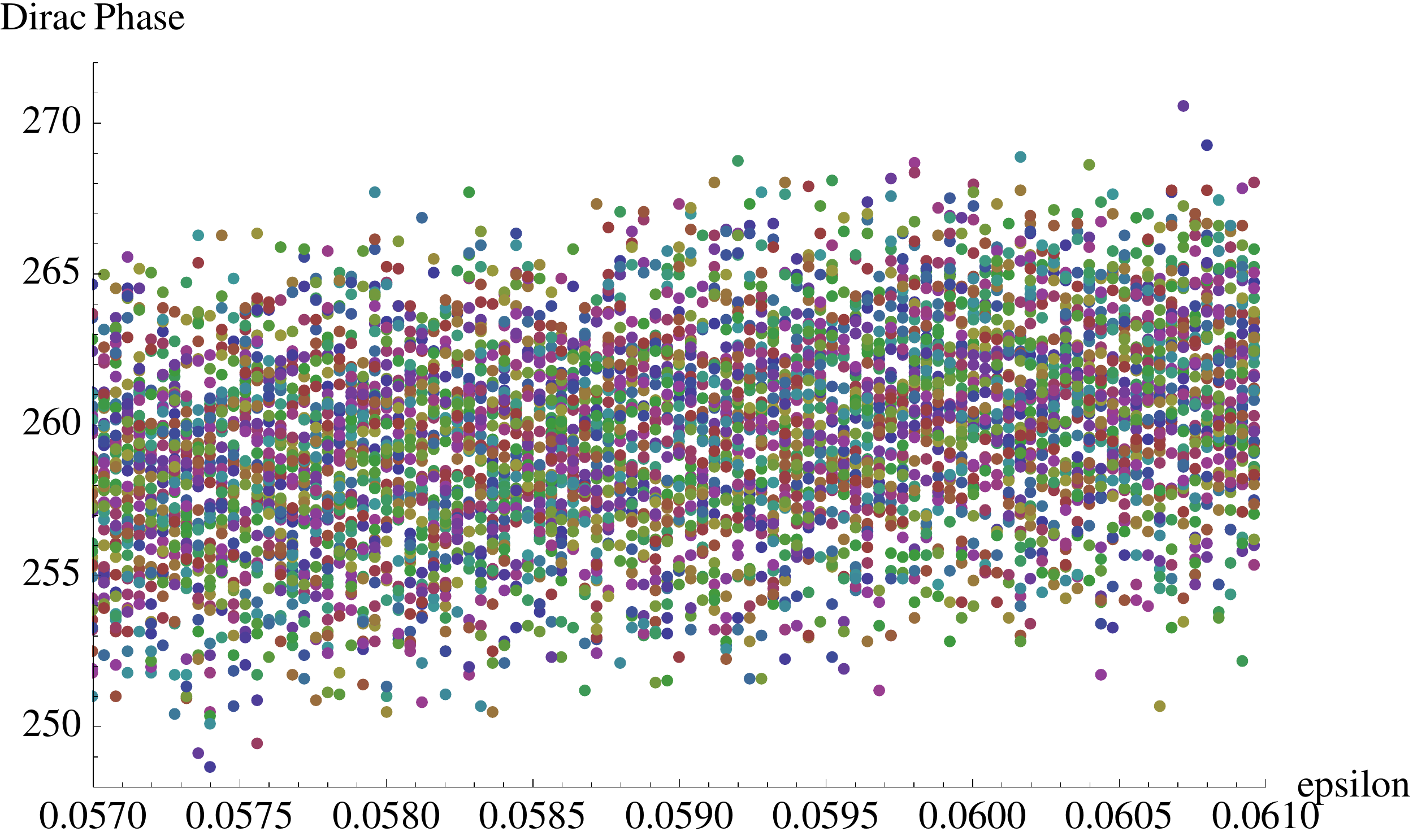}
    \caption{Left panel shows a zoom-in of the predictions for the atmospheric angle $\theta^l_{23}$ (upper) and
    solar angle $\theta^l_{12}$ (lower) in the presence of HO corrections. Right panel shows a zoom-in of the predictions for the Dirac oscillation phase $\delta^l$ (in degrees)
    in the presence of HO corrections.
    The predictions are presented as a function of 
			$\epsilon_{\nu}=m_b/m_a$ and hence $m_2/m_3$,
			for $m_1=0.3$ meV and $m_2=50$ meV.
			Note that these predictions assume $\eta = 2\pi/5$.
			The predictions are obtained numerically using the Mixing Parameter Tools (MPT)
package based on \cite{Antusch:2005gp}, taking into account the different conventions.
} \label{predictions}
\vspace*{-2mm}
\end{figure}

\begin{figure}
\centering
\includegraphics[width=0.49\textwidth]{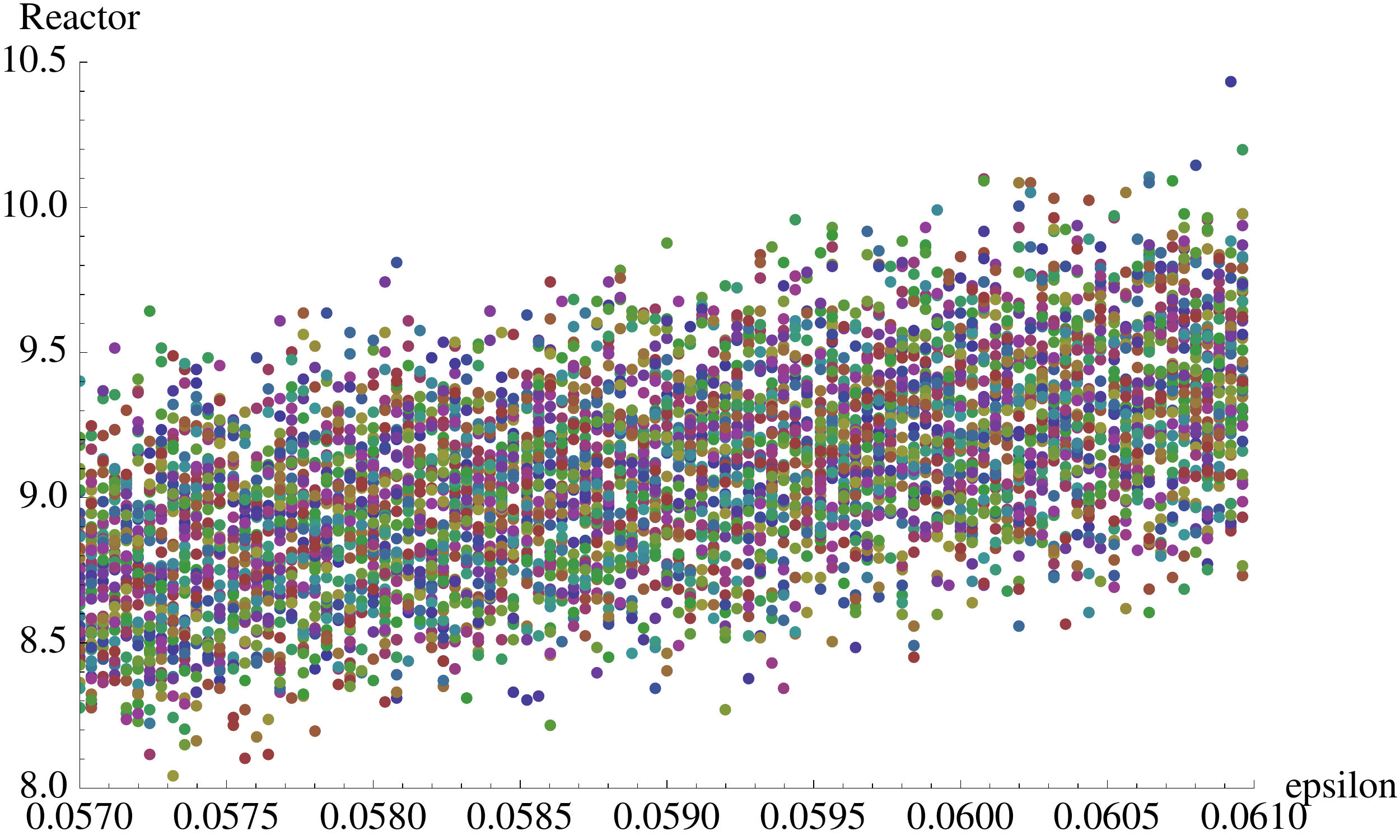}
\includegraphics[width=0.49\textwidth]{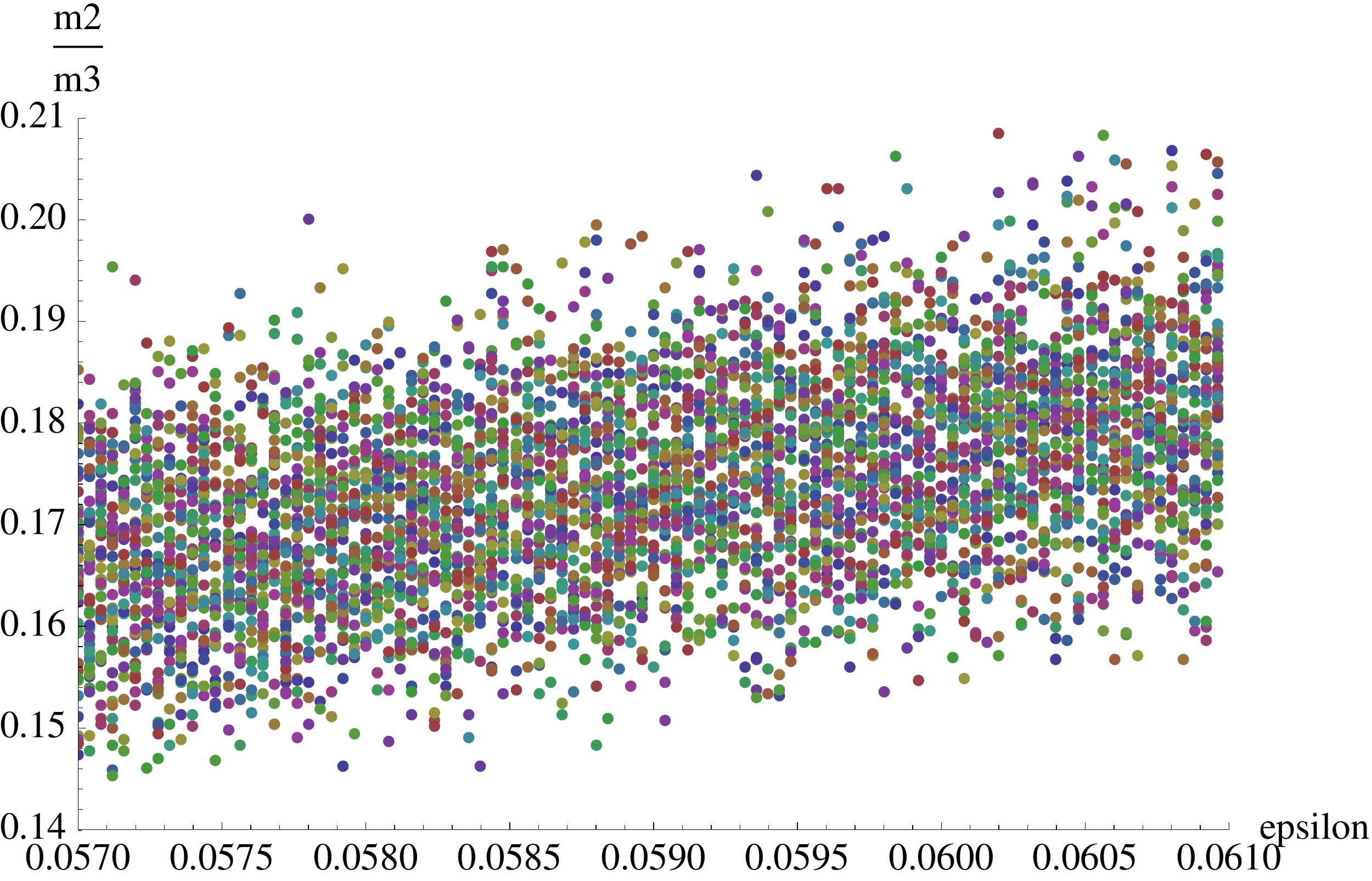}
    \caption{Left panel shows a zoom-in of the predictions for the reactor angle $\theta^l_{13}$ 
    as a function of $\epsilon_{\nu}=m_b/m_a$, in the presence of HO corrections. Right panel the predictions for the neutrino mass ratio $m_2/m_3$ as a function of $\epsilon_{\nu}=m_b/m_a$, in the presence of HO corrections.
    Taken together, these plots show how the reactor angle prediction increases with $m_2/m_3$.
    The predictions are presented
			for $m_1=0.3$ meV and $m_2=50$ meV.
			Note that these predictions assume $\eta = 2\pi/5$.
			The predictions are obtained numerically using the Mixing Parameter Tools (MPT)
package based on \cite{Antusch:2005gp}, taking into account the different conventions.
} \label{reactor}
\vspace*{-2mm}
\end{figure}

\begin{equation} \label{YnuHO}
Y^{\nu} =  \begin{pmatrix}  \varepsilon_{11}\epsilon^2 & b\epsilon (1+\varepsilon_{12})  & \varepsilon_{13}\\ 
a\epsilon^2(1+\varepsilon_{21}) & 4b\epsilon (1+\varepsilon_{22}) &  \varepsilon_{23}\\  
a\epsilon^2(1+\varepsilon_{31}) & 2b\epsilon (1+\varepsilon_{32}) & c/3(1+ \varepsilon_{33})\end{pmatrix}.
\end{equation}
The corrections in the third column are not important for lepton mixing, but we include them in the scans.
Due to Clebsch factors in the HO corrections, we consider the neutrino corrections to be
independent of the up quark corrections, but of the same order of magnitude.

 In Fig.\ref{overview} we show the predictions for the PMNS atmospheric and solar angles and all the phases, obtained from the Yukawa matrix
 in Eq.\ref{YnuHO} as a function of the ratio $\epsilon_{\nu}=m_b/m_a$ where we have implemented the see-saw mechanism leading to a light effective Majorana neutrino mass matrix as in Eq.\ref{seesaw3}, but involving the
 HO corrections $ \varepsilon_{ij}$.
 In Fig.\ref{predictions} we show a blow-up of the atmospheric and solar angle predictions, together with the
 Dirac CP violating oscillation phase. 
 The reactor angle has a stronger correlation with $\epsilon_{\nu}=m_b/m_a$ and hence $m_2/m_3$ as shown in Fig.~\ref{reactor}.
 These results may be compared to the LO predictions shown in Table~\ref{predictions142}.
In Figs.~\ref{overview},\ref{predictions},\ref{reactor} we have randomly scanned 
over the independent (uncorrelated) complex parameters $ \varepsilon_{ij}$ which are allowed to 
 take complex values with real and imaginary parts randomly chosen between $-0.02$ and $0.02$, limiting the modulus
 to be less than about $0.03$.
In the scans we have assumed that the corrections $ \varepsilon_{11}$ and  $ \varepsilon_{13}$
are smaller by a factor of 1/4 than the other corrections, 
$|\frac{ \varepsilon_{11}}{\varepsilon_{21}}|\sim 1/4$ and  $| \frac{\varepsilon_{13}}{ \varepsilon_{23}}|\sim 1/4$,
due to the pattern of alignments in Eq.\ref{phiu}. This assumption is consistent with the fact that in the quark sector we require $|\frac{\varepsilon_{13}}{\varepsilon_{23}}|\sim 0.22$.

From the plots in Figs.~\ref{overview},\ref{predictions},\ref{reactor} we estimate the approximate one sigma ranges 
for the theoretical predictions of the atmospheric and solar angles of 
 $\theta^l_{23}=40^\circ \pm1^\circ$, $\theta^l_{12}=34^\circ \pm1^\circ$,
 with a reactor angle $\theta^l_{13}=9.0^\circ \pm 0.5^\circ$, correlated with $m_2/m_3$.
We also predict the Dirac CP violating oscillation phase to be $\delta^l=260^\circ \pm 5^\circ$.
The predictions of the tetra-model of a normal hierarchy of neutrino masses
with an atmospheric angle in the first octant will be tested quite soon.
We emphasise that the above errors arise from the same higher order corrections
which are solely responsible for the small quark mixing angles.
This gives a handle on the size of the irreducible theoretical error that must be included in the leptonic
predictions. In addition, we expect additional corrections of possibly similar magnitude arising from
renormalisation group (RG) running \cite{Boudjemaa:2008jf}
and canonical normalisation corrections \cite{Antusch:2007ib}.
For example in SUSY GUT models and light sequential dominance, similar to the case here,  
the RG corrections for high $\tan \beta\sim 50$ are \cite{Boudjemaa:2008jf}: 
$\Delta \theta^l_{23}\sim +1^\circ$,
$\Delta \theta^l_{12}\sim +0.4^\circ$, $\Delta \theta^l_{13}\sim - 0.1^\circ$,
where the ``$+$'' sign means that the value increases in running from the GUT scale to low energy,
while for low $\tan \beta \lsim 10$ the RG corrections are negligible.
Clearly the RG corrections provide additional shifts in the central values of the predicted angle, but the shifts 
lie within the errors quoted above arising from HO corrections,
and for low $\tan \beta \lsim 10$ such RG shifts are negligible.
Clearly knowledge of the error in the leptonic predictions is crucial since such predictions
will be subject to intense experimental scrutiny over the coming years
and will serve to test the tetra-model.

\section{Conclusions}
\label{conclusions}
 
In this paper we have proposed a tetra-model of quark and lepton mixing
based on tetrahedral $A_4$ family symmetry and the tetra-colour Pati-Salam group $SU(4)_{PS}$
together with $SU(2)_L \times U(1)_R$ and the tetra-vacuum alignment $(1,4,2)$.
Leptonic mixing and CP violation is fully predicted 
at leading order, as a consequence
of the vacuum alignment. In addition, a 
Cabibbo angle
$\theta_C\approx 1/4$ emerges from the tetra-alignment $(1,4,2)$, 
which appears in the second column common to the neutrino and up quark Yukawa matrices,
providing the Cabibbo connection between quark and lepton mixing.

Due to the requirement of having diagonal down and charged lepton Yukawa matrices, with all quark and lepton
mixing originating from the up and neutrino Yukawa matrices, we do not impose the $SU(2)_R$ gauge group,
only its diagonal subgroup $U(1)_R$ where $R=T_{3R}$, the third (diagonal) $SU(2)_R$ generator.
For this reason, and also the fact that the left-handed and right-handed quarks and leptons transform differently
under $A_4$, the tetra-model cannot (easily) be embedded into $SO(10)$ at the field theory level.
However, we speculate that it may 
be possible to obtain the tetra-model directly from string theories such as heterotic string theory, 
F-theory or M-theory in which $SO(10)$ is present in extra dimensions.
Supersymmetry (SUSY) is motivated by both string theory and the vacuum alignment mechanism,
as well as gauge coupling unification and dark matter, however there are no other compelling 
reasons why the tetra-model could not be formulated as a non-SUSY model.

The leading order Yukawa matrices and Majorana neutrino mass matrix have a remarkably simple
form as shown in Eqs.~\ref{Yed2}, \ref{Yunu2}, involving only nine real parameters, namely $y_d,y_s,y_b$, $|a|,|b|,|c|$ and $M_1,M_2,M_3$ (where we may absorb the powers of $\epsilon$) leading to the predictions summarised in section~\ref{leading}.
The charged lepton masses are related to down quark masses by GJ relations.
Dirac neutrino masses are equal to up quark masses, up to Clebsch-Gordan coefficients, while 
Majorana neutrino masses are proportional to the squares up type quark masses,
giving an SO(10)-like pattern.
However in our model the strong hierarchies naturally cancel in the see-saw mechanism, leading to a normal neutrino mass hierarchy. Lepton mixing angles and all CP phases are predicted as a function of the neutrino mass ratio $m_2/m_3$, with the angles being affected by about one degree from the results in the previous two right-handed neutrino model due to the non-zero lightest neutrino mass.

At leading order, the model provides a good description of the twelve fermion masses (including the light three neutrino masses) as well as the six PMNS parameters and the Cabibbo angle: a total of nineteen physical observables from nine input parameters, which is ten fewer parameters than in the SM. The ten leading order predictions include the six PMNS parameters, three charged lepton-down quark mass relations and the Cabibbo angle.
The lepton mixing angles and the Cabibbo angle can be understood as 
arising from the vacuum alignment of the $A_4$ family symmetry breaking flavons. Leptonic CP violation
arises from a relative phase of $-4\pi/5$ between 
$Z_5$ breaking flavon VEVs 
which appear in the construction of the neutrino mass matrix in Eq.~\ref{seesaw2}.
This phase
is essential in obtaining the correct leptonic mixing angles although it plays no simple role
in the quark mixing angles since 
the Cabibbo angle is independent of this phase and the other quark
mixing angles and CP phase are zero at the leading order.
The remaining 3 parameters, namely 
the two small quark mixing angles $\theta^q_{13}$, $\theta^q_{23}$ associated with $V_{ub}$, $V_{cb}$
and the CP violating phase $\delta^q$, are zero at leading order and 
originate from a large number of higher order operators.
The higher order operators 
also correct Cabibbo angle and PMNS parameters, leading to some theoretical error or ``noise''
in the leading order predictions.

One of the main successes of the
tetra-model is that it provides an explanation for why the Cabibbo angle and 
lepton mixing angles take the ``large'' values that they do, as a result of the vacuum alignment of the
$A_4$ symmetry breaking flavons, while the remaining
quark mixing angles are ``small'' since they are zero at leading order and become non-zero due to 
higher order corrections. The model provides an explanation for the 
size of the Cabibbo angle and its role as the link between quark and lepton mixing
(the Cabibbo connection).
The higher order corrections also affect the Cabibbo angle and PMNS parameters,
providing a source of theoretical error or ``noise'' which blurs the leading order predictions.
In the case of the Cabibbo angle, such ``noise'' is in fact necessary in order to bring the leading order
prediction of $\theta_C\approx 14^\circ $ into precise agreement with experiment
where $\theta_C\approx 13^\circ $.
In the case of the lepton mixing angles, the ``noise'' provides a guide to the experimental accuracy 
required in order to test the model. 

It is worth briefly discussing the experimental prospects
for testing the predictions of the tetra-model.
The ``binary'' predictions of a normal neutrino mass hierarchy and an atmospheric angle in the first octant
will both be tested over the next few years by current and planned neutrino experiments
such as SuperKamiokande, T2K, NO$\nu$A and PINGU \cite{Winter:2013ema}. 
The Daya Bay II reactor experiment, including the 
short baseline detectors \cite{Zhan:2013cxa}, will also test the neutrino mass hierarchy and in addition
measure the reactor and solar angle to high accuracy,
enabling precision tests of the predictions $\theta^l_{13}=9.0^\circ \pm 0.5^\circ$
and $\theta^l_{12}=34^\circ \pm1^\circ$.
In the longer term, superbeam and neutrino factory proposals such as WBB and LENF 
\cite{Ballett:2013wya} would measure 
the atmospheric mixing angle to high accuracy, confronting the prediction $\theta^l_{23}=40^\circ \pm1^\circ$,
and ultimately testing the
prediction of the Dirac CP violating oscillation phase $\delta^l=260^\circ \pm 5^\circ$.

In conclusion, the tetra-model is a robust theory of flavour based on quark-lepton-family unification. It solves many
of the flavour puzzles, halving the number of parameters in the flavour sector of the SM.
At leading order, the tetra-model remarkably gives 
ten predictions, somewhat blurred by the higher order corrections as we have discussed.
It also qualitatively explains the smallness of the quark mixing angles compared to lepton mixing angles, with the former being zero at leading order, apart from
the Cabibbo angle which is given by $\theta_C\approx 1/4$,
due to the tetra-vacuum alignment $(1,4,2)$,
providing the Cabibbo connection between quark and lepton mixing.
The tetra-model involves an $SO(10)$-like pattern of Dirac and heavy right-handed neutrino masses,
with the strong up-type quark mass hierarchy cancelling in the see-saw mechanism, leading to a 
relatively mild normal hierarchy of neutrino masses.
The tetra-model yields fairly accurate predictions for all six PMNS mixing parameters 
(three angles as well as three phases) and predicts a normal neutrino mass hierarchy with the 
atmospheric angle in the first octant.
It will be decisively tested over the coming years by presently running and future neutrino 
experiments. 

\section*{Acknowledgements}
SFK would like to thank Pasquale Di Bari and Claudia Hagedorn for discussions.
SFK also acknowledges partial support 
from the STFC Consolidated ST/J000396/1 and EU ITN grants UNILHC 237920 and INVISIBLES 289442 .

\end{document}